\documentclass[aps,pre,reprint,superscriptaddress,longbibliography,floatfix,nofootinbib]{revtex4-2}

\usepackage[T1]{fontenc}
\usepackage[utf8]{inputenc}
\usepackage{amsmath,amssymb,amsfonts,bm}
\usepackage{graphicx}
\usepackage{booktabs}
\usepackage{xcolor}
\usepackage[colorlinks=true,linkcolor=blue,citecolor=blue,urlcolor=blue]{hyperref}
\usepackage[capitalise,nameinlink]{cleveref}

\graphicspath{{figures/}}
\allowdisplaybreaks

\definecolor{revisionblue}{rgb}{0,0,0}

\DeclareRobustCommand{\rev}[1]{{\color{revisionblue}#1}}

\newcommand{\pqnum}[1]{\left[#1\right]_{p,q}}
\newcommand{\dd}{\mathop{}\!\mathrm{d}}
\newcommand{\Id}{\mathbb{I}}
\newcommand{\Tr}{\operatorname{Tr}}
\newcommand{\QM}{Q_{\mathrm{M}}}
\newcommand{\QT}{Q_{\mathrm{T}}}
\newcommand{\Wcyc}{W_{\mathrm{cyc}}}
\newcommand{\Wout}{W_{\mathrm{out}}}
\newcommand{\sigmat}{\widetilde{\sigma}}

\begin{document}

\title{\texorpdfstring{\textcolor{revisionblue}{Thermodynamic performance of a measurement-driven quantum engine with a two-parameter $(p,q)$-deformed harmonic oscillator}}{Thermodynamic performance of a measurement-driven quantum engine with a two-parameter (p,q)-deformed harmonic oscillator}}

\author{Tunde Joseph Osunmusanmi}
\email{Tunde.josunmusanmi@ku.ac.ae}
\affiliation{College of Computing and Mathematical Sciences, Department of Applied Mathematics and Sciences, Khalifa University of Science and Technology, 127788 Abu Dhabi, United Arab Emirates}

\author{Berihu Teklu}
\email{berihu.gebrehiwot@ku.ac.ae}
\affiliation{College of Computing and Mathematical Sciences, Department of Applied Mathematics and Sciences, Khalifa University of Science and Technology, 127788 Abu Dhabi, United Arab Emirates}
\affiliation{KU Research Center for Intelligent Computing, Networks and Sensing (ICONS), Khalifa University of Science and Technology, 127788, Abu Dhabi, United Arab Emirates}

\date{\today}

\begin{abstract}
We address a single-bath measurement-driven engine with a two-parameter $(p,q)$-deformed oscillator spectrum. A nonselective Gaussian quadrature measurement supplies energy between two population-preserving work strokes. We evaluate its finite-dimensional transition matrix without outcome quadrature and derive the leading weak-measurement heat. Uniform spectral scaling gives the Otto efficiency exactly; deformation changes the relative gaps and the conversion of measurement energy into work. The infinite quadrature is essentially self-adjoint for $\max(p,q)\le1$; above this range a self-adjoint extension must be specified.
We instead choose an explicit $N=40$ effective matrix model. At $\beta\hbar\omega_i=1$, fixed physical resolution $\sigma=4$, and $\omega_f/\omega_i=2$, the largest reported efficiency in $0.7\le p_f,q_f\le1.2$ is approximately $0.782$, compared with the undeformed value $0.5$. Independent dimension checks reproduce these values but do not establish an infinite-dimensional limit in the enhanced regime. Comparisons at selected matched resources and at fixed mean deformation reveal dependence on spectral shape beyond first order. The working-medium efficiency excludes measurement-apparatus and control implementation costs.

\end{abstract}

\maketitle

\section{Introduction}
\label{sec:introduction}

Quantum heat-engine models date back to the three-level maser~\cite{Scovil1959PRL}.
The quantum Otto cycle for few-level systems and harmonic oscillators is a standard benchmark in quantum thermodynamics~\cite{Kieu2004PRL,Rezek2006NJP,Quan2007PRE,Kosloff2017Entropy}.
It has become a broad framework for energy conversion in nanoscale and coherently controlled devices~\cite{Kosloff2013Entropy,Vinjanampathy2016ContempPhys}.
Theoretical and experimental works have extended it beyond idealized two-bath cycles to machines driven by engineered squeezed reservoirs~\cite{Rossnagel2014PRL},
coherence~\cite{Scully2003Science,Uzdin2015PRX}, correlations~\cite{PerarnauLlobet2015PRX},
measurements~\cite{Yi2017PRE,Elouard2018PRL,Chand18,Chand,Dieguez,Purkait}, many-body
effects~\cite{Jaramillo2016NJP}, criticality~\cite{Campisi2016NatCommun,Fogarty2021QST},
prethermalization~\cite{Brollo2025NatCommun}, and explicitly time-dependent control,
including shortcuts to adiabaticity~\cite{delCampo2014SciRep}.
Experiments on single trapped ions~\cite{Rossnagel2016Science}, atomic
collisions~\cite{Bouton2021NatCommun}, many-body quantum
gases~\cite{Koch2023Nature}, exceptional-point dynamics~\cite{Zhang2022NatCommun,Bu2023PRL},
noise-assisted superconducting thermal machines~\cite{Sundelin2026NatCommun}, and
dissipation-engineered superconducting circuits~\cite{Uusnakki2026NatCommun} show that
quantum engines are not only conceptual models but platforms where energetic exchanges
can be monitored and optimized at the level of discrete transitions.
Hence, quantum thermal machines offer a controlled setting in which the validity of
definitions of work, heat, entropy production and fluctuations in microscopic working
media can be tested. The structure of the spectrum and the set of accessible conserved
quantities can modify the performance of the engine beyond simple few-level models, as
shown in many-body, critical, and prethermal working media
\cite{Fogarty2021QST,Brollo2025NatCommun,Campbell2026QST}. In a recent
study~\cite{Yi2017PRE}, it was shown that in a measurement-driven engine, the
energetic input is not provided by a usual hot reservoir but is instead injected by
the measurement apparatus through the state transformation of the measurement channel
(measurement backaction). The corresponding heat and work statistics were analyzed in
detail in~\cite{Ding2018PRE}.

Related work has demonstrated the extraction of work from measurement records via
feedback and Maxwell-demon protocols \cite{Elouard2017PRL}, the powering of efficient
engines via position-resolving measurements \cite{Elouard2018PRL}, the limitations of
the accessible power-efficiency regime by pointer measurements \cite{Seah2020PRL}, the
use of local measurements and entanglement as fuel \cite{Bresque2021PRL}, and the use
of incompatible measurements as fuel for oscillator engines
\cite{Manikandan2022PRE}. Measurement-control variants, such as quantum Zeno strokes
that replace slow adiabatic transformations with frequent selective measurements
\cite{Barontini2025PRL}, provide yet another route to measurement-assisted finite-time
engines and heat pumps. Furthermore, extensions to multilevel systems, finite-time
cycles, ancilla-assisted measurement architectures, explicit measurement costs,
entangling measurements, ergotropy-assisted operation, finite-duration measurement
effects, measurement-fueled energetics in quantum-computer and superconducting-circuit
settings, information-engine performance bounds, and work/heat fluctuations in
measurement-based engines and batteries have been explored
~\cite{Anka2021PRE,Wang2024PRE,Rathnakaran2025PRE,Perna2024PRE,
Mayo2026PRE,Jakhar2026PRE,YiKim2026PRA,Kirchberg2025PRA,
ElMakouri2026APSOpenSci,Solfanelli2021PRXQuantum,Dassonneville2026PRResearch}.

Most oscillator-based measurement engines use the standard harmonic spectrum, in which
adiabatic control changes the level spacing only through the oscillator frequency.
The present work addresses a complementary control mechanism: spectral engineering of
the working medium. A two-parameter $(p,q)$-deformed oscillator provides a richer
spectral structure, in which the level spacings depend nonlinearly on the excitation
number and on the deformation pair
$(p,q)$~\cite{Chakrabarti1991JPhysA,Jagannathan2006arxiv,
Lorek1997ZPhysC,Algin2024EPJP,Sargolzaeipor2019Pramana}. Related work has shown that
$q$-deformation of the working substance can itself act as a thermodynamic control
parameter in a quantum Otto cycle~\cite{Ozaydin2023PRE}. More recently,
measurement-controlled quantum heat engines with a $q$-deformed harmonic-oscillator
working substance have been proposed~\cite{TONG2026131619}. These studies motivate
asking whether spectral deformation can also modify the conversion of measurement
backaction into useful work in a single-bath measurement-driven cycle, and whether a
second deformation parameter adds anything beyond the one-parameter case.

Our primary aim is to determine whether deformation-induced spectral nonlinearity
modifies the performance of a single-bath measurement engine beyond a simple rescaling
of the oscillator frequency. This distinction is important because widening all energy
gaps by a common factor can increase energetic quantities without providing a
meaningful thermodynamic advantage. We make this distinction precise. We show that,
for any uniformly scaled spectrum, the measurement-driven cycle yields exactly the
Otto efficiency $1-\omega_i/\omega_f$, whatever the measurement strength, the
temperature, or the source of the energy input. A departure from the Otto value is
therefore a direct signature of the change in relative level spacing produced by the
deformation, and we use the undeformed oscillator, whose performance is known in
closed form, as the baseline. We study the working-medium efficiency,
\begin{equation}
 \eta=\frac{\Wout}{\QM}=\frac{-\Wcyc}{\QM},
 \label{eq:intro_efficiency}
\end{equation}
where $\QM$ is the energy injected into the oscillator by the nonselective measurement.
Moreover, $\Wcyc$ is the net work done on the working medium. This is not the complete
device-level efficiency of an autonomous machine, which accounts for the energy cost
of implementing, recording, and resetting the measurement apparatus. In limited
circumstances, idealized projective measurements can demand unbounded resources
\cite{Guryanova2020Quantum}, while realistic measurement models impose additional
energetic, temporal, and dynamical limitations
\cite{Latune2025Quantum, Perna2024PRE, Kirchberg2025PRA, YiKim2026PRA}. We
therefore call $\eta$ the working-medium efficiency: the ratio of the useful work output
to the energy injected into the working medium by the measurement, excluding the
energetic cost of implementing, recording, and resetting the measurement apparatus. It
should not be confused with a full device-level efficiency.

We formulate a four-stroke cycle in which the working medium is initially thermalized
with a single bath at inverse temperature $\beta$. Stroke I is an adiabatic
change of the control vector $\lambda=(\omega,p,q)$ from
$\lambda_i=(\omega_i,p_i,q_i)$ to $\lambda_f=(\omega_f,p_f,q_f)$, preserving level
populations while changing the spectrum. Stroke II is a nonselective Gaussian
measurement of the final deformed position quadrature. Stroke III returns the
Hamiltonian parameters to their initial values adiabatically, and stroke IV
rethermalizes the oscillator with the original bath. We derive the measurement-induced
heat, both adiabatic work contributions, the heat exchanged during thermalization, and
the working-medium efficiency under a single sign convention, and identify the physically
admissible engine regime by imposing $\QM>0$, $\Wcyc<0$, $\QT<0$, and $0<\eta<1$.

Our contribution is a finite-model analysis of how spectral shape affects the
conversion of nonselective measurement backaction into work. We distinguish uniform
scaling from changes in relative gaps, identify the first-order symmetric deformation
and higher-order asymmetry, and separate dimension sensitivity from convergence to an
infinite oscillator. The larger work and efficiency values below concern the specified
finite model and selected comparison constraints; they do not establish a universal
resource advantage or a physical implementation of deformation control.

The paper is organized as follows. Section~\ref{sec:model} defines the
$(p,q)$-deformed oscillator. Section~\ref{sec:cycle} derives the measurement map, the
cycle energetics, and the exact Otto result for uniformly scaled spectra.
Section~\ref{sec:numerics} discusses the spectral validity of the model, the exact
evaluation of the transition matrix, the convergence criteria, and the numerical
tests. Section~\ref{sec:results} presents the deformation dependence of the
measurement heat, the adiabatic work, the net work, and the efficiency.
Section~\ref{sec:conclusion} summarizes the conclusions and limitations.

\section{Two-parameter deformed oscillator}
\label{sec:model}

We consider a working medium described by a two-parameter $(p,q)$-deformed harmonic oscillator with $p,q>0$. The annihilation and creation operators \(a_{p,q}\) and $a_{p,q}^{\dagger}$ obey
\cite{Chakrabarti1991JPhysA,Jagannathan2006arxiv,Bonatsos1999,
Bera2006,Lorek1997ZPhysC,Boumali2017,NaseriKarimvand2022,
Algin2024EPJP,Sargolzaeipor2019Pramana}
\begin{equation}
\begin{aligned}
a_{p,q}a_{p,q}^{\dagger}
 - q\,a_{p,q}^{\dagger}a_{p,q}
&= p^{N}, \\
[N,a_{p,q}^{\dagger}]
&= a_{p,q}^{\dagger}, \\
[N,a_{p,q}]
&= -a_{p,q}.
\end{aligned}
\label{eq:pq_algebra}
\end{equation}
where $N$ is the number operator. The corresponding $(p,q)$-basic number is
\begin{equation}
 \pqnum{n}=\frac{p^n-q^n}{p-q},
 \qquad p\neq q.
 \label{eq:pq_basic}
\end{equation}
For integer $n\geq 1$, this can be written as
\begin{equation}
 \pqnum{n}=\sum_{r=0}^{n-1}p^{n-1-r}q^r,
 \label{eq:pq_sum}
\end{equation}
which shows that $\pqnum{n}\to n$ in the undeformed limit $p,q\to 1$. The continuous diagonal limit is
\begin{equation}
 [n]_{s,s}=n s^{n-1},
 \qquad s=p=q.
 \label{eq:diagonal_limit}
\end{equation}
The deformed Fock states satisfy the following conditions
\begin{align}
 N|n\rangle &= n|n\rangle,\label{eq:number_action}\\
 a_{p,q}|n\rangle &= \sqrt{\pqnum{n}}\,|n-1\rangle,\label{eq:lowering_action}\\
 a_{p,q}^{\dagger}|n\rangle &= \sqrt{\pqnum{n+1}}\,|n+1\rangle,\label{eq:raising_action}
\end{align}
with $a_{p,q}^{\dagger}a_{p,q}=\pqnum{N}$ and $a_{p,q}a_{p,q}^{\dagger}=\pqnum{N+1}$. The number states may be generated from the vacuum by
\begin{equation}
 |n\rangle=\frac{(a_{p,q}^{\dagger})^n}{\sqrt{\pqnum{n}!}}|0\rangle,
 \qquad
 \pqnum{n}!=\prod_{k=1}^{n}\pqnum{k}.
 \label{eq:fock_generation}
\end{equation}

The Hamiltonian is taken as
\begin{equation}
 H(\omega,p,q)=\frac{\hbar\omega}{2}\left(a_{p,q}^{\dagger}a_{p,q}+a_{p,q}a_{p,q}^{\dagger}\right),
 \label{eq:hamiltonian}
\end{equation}
with eigenvalues
\begin{equation}
 E_n(\omega,p,q)=\frac{\hbar\omega}{2}\left[\pqnum{n}+\pqnum{n+1}\right],
 \quad n=0,1,2,\ldots .
 \label{eq:energy_spectrum}
\end{equation}
The corresponding deformed quadratures are
\begin{align}
 \hat{x}_{p,q} &= \sqrt{\frac{\hbar}{2m\omega}}\left(a_{p,q}+a_{p,q}^{\dagger}\right),\label{eq:position}\\
 \hat{\pi}_{p,q} &= i\sqrt{\frac{m\hbar\omega}{2}}\left(a_{p,q}^{\dagger}-a_{p,q}\right),\label{eq:momentum}
\end{align}
where $\hat{\pi}_{p,q}$ denotes the momentum operator and avoids notational confusion with the deformation parameter $p$. In terms of these quadratures, the Hamiltonian can be written as
\begin{equation}
 H(\omega,p,q)=\frac{\hat{\pi}_{p,q}^2}{2m}+\frac{1}{2}m\omega^2\hat{x}_{p,q}^2,
 \label{eq:quadrature_hamiltonian}
\end{equation}
under the same deformed algebra. In the undeformed limit $p,q\to 1$, Eq.~\eqref{eq:energy_spectrum} reduces to the ordinary harmonic oscillator spectrum $E_n=\hbar\omega(n+1/2)$. Equation~\eqref{eq:quadrature_hamiltonian} is an algebraic identity, not a statement that $\hat{x}_{p,q}$ and $\hat{\pi}_{p,q}$ are canonical observables. From Eqs.~\eqref{eq:position} and \eqref{eq:momentum} one finds $[\hat{x}_{p,q},\hat{\pi}_{p,q}]=i\hbar\left([N+1]_{p,q}-[N]_{p,q}\right)$, which reduces to the canonical value $i\hbar$ only for $p=q=1$. In this work, $\hat{x}_{p,q}$ is the observable accessed by the measurement stroke; its mathematical properties are discussed in Sec.~\ref{sec:numerics} and Appendix~\ref{subsec:finite_truncation}.

The adjacent-level spacing captures the physical role of the deformation
\begin{equation}
 \Delta_n(\omega,p,q)=E_{n+1}(\omega,p,q)-E_n(\omega,p,q).
 \label{eq:level_spacing}
\end{equation}
For $p=q=1$, $\Delta_n=\hbar\omega$ is independent of $n$. For $p\neq 1$ or $q\neq 1$, the level spacing is generally excitation-dependent. Measurement-induced transitions, therefore, carry deformation-dependent energy differences, and adiabatic work strokes depend on how the full spectrum changes when $(\omega,p,q)$ is varied. A small-deformation expansion about $p=q=1$ is given in Appendix \ref{app:perturbative_expansion}; it shows that the leading spectral correction is governed by the symmetric combination $\delta=(p+q)/2-1$ and is quadratic in the excitation number.

Since $[n]_{p,q}=[n]_{q,p}$, the two parameters do not have separate operational meanings. The physically distinct combinations are the mean deformation $\delta$, which sets the leading Kerr-type nonlinearity of the spectrum, and the asymmetry $\gamma=(p-q)/2$, which enters at second order and, through $\max(p,q)$, controls the growth of the spectrum at large $n$. We treat $(\omega,p,q)$ as control parameters of a Hamiltonian that is a function of the number operator alone. Operationally, varying them amounts to tuning the energies of the oscillator levels while keeping the eigenbasis fixed, and the work done by the control is the resulting change of energy at fixed populations. The independent tunability of these levels and access to the deformed quadrature
are assumptions of the effective model. We do not supply a microscopic control
Hamiltonian, an experimental deformation protocol, or a detector-coupling realization.
The energetic cost of these controls is not included.

\section{Cycle, measurement map, and thermodynamic quantities}
\label{sec:cycle}

\begin{figure}[tbp]
  \centering
  \includegraphics[width=\columnwidth]{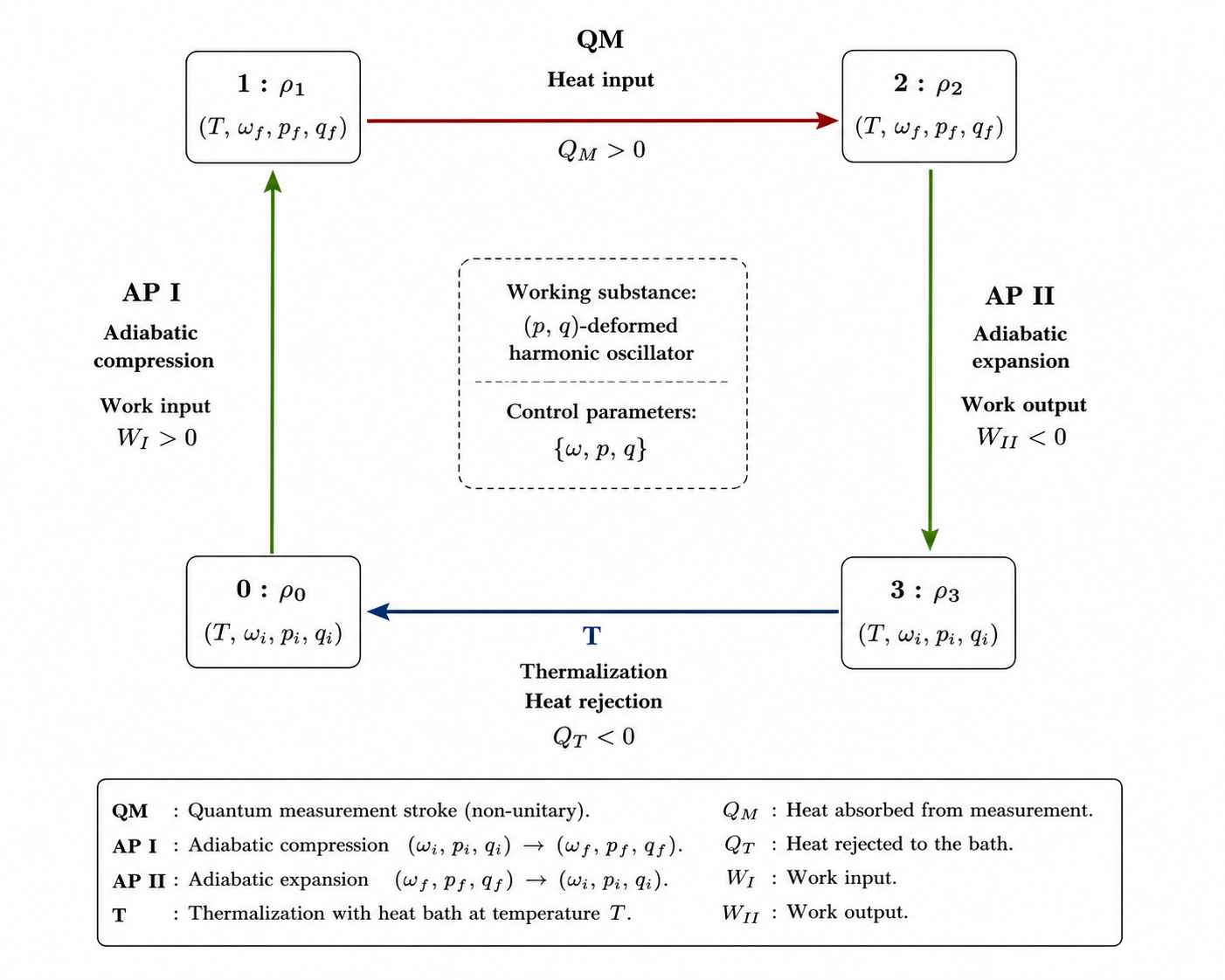}
 \caption{Four-stroke measurement-driven engine. The system starts in a Gibbs state of the initial Hamiltonian $H_i$. Stroke I is an adiabatic change of the control parameters from $\lambda_i$ to $\lambda_f$, preserving level populations while changing the spectrum. Stroke II is a nonselective Gaussian measurement of the final deformed position quadrature, which injects energy $\QM$. Stroke III returns the Hamiltonian parameters to $\lambda_i$, and stroke IV rethermalizes the oscillator with the original bath. Work is defined as work done on the system; useful work output corresponds to $\Wcyc<0$. The external controls are the parameters $\lambda=(\omega,p,q)$, varied during strokes I and III; the measurement apparatus is treated as an external, unmodeled resource whose energetic cost is not included.}
 \label{fig:cycle}
\end{figure}

The four-stroke cycle is shown schematically in Fig.~\ref{fig:cycle}. We denote the initial and final Hamiltonian parameters by
\begin{equation}
 \lambda_i=(\omega_i,p_i,q_i),
 \qquad
 \lambda_f=(\omega_f,p_f,q_f),
 \label{eq:lambda_definitions}
\end{equation}
with $H_i=H(\lambda_i)$ and $H_f=H(\lambda_f)$. The corresponding energy eigenvalues are $E_n^{(i)}=E_n(\lambda_i)$ and $E_n^{(f)}=E_n(\lambda_f)$.

\subsection{Initial thermal state and first adiabatic stroke}
\label{subsec:first_stroke}

The engine begins in a Gibbs state at inverse temperature $\beta=(k_{\mathrm B}T)^{-1}$,
\begin{equation}
 \rho_0=\sum_{n=0}^{\infty}P_n^{(i)}|n;i\rangle\langle n;i|,
 \qquad
 P_n^{(i)}=\frac{e^{-\beta E_n^{(i)}}}{Z_i},
 \label{eq:gibbs_state}
\end{equation}
where
\begin{equation}
 Z_i=\sum_{n=0}^{\infty}e^{-\beta E_n^{(i)}}.
 \label{eq:partition_function}
\end{equation}
In numerical work, the sums are restricted to $n\le N-1$; the role of the dimension
$N$ is discussed in Sec.~\ref{sec:numerics}. The truncated probabilities
$P_n^{(i,N)}=e^{-\beta E_n^{(i)}}/Z_i^{(N)}$, with $Z_i^{(N)}$ the sum in
Eq.~\eqref{eq:partition_function} restricted to $n\le N-1$, differ from the exact ones by the
discarded weight $\Pr(n\ge N)=e^{-N\beta\hbar\omega_i}$. At the parameters used in
this work ($\beta\hbar\omega_i=1$, $N=40$) this weight is $4\times10^{-18}$, and the
errors in weighted energy moments require separate bounds and cannot be inferred
from this probability alone. For an incoming harmonic Gibbs state, with
$r=e^{-\beta\hbar\omega_i}$, the omitted unnormalized first energy moment is
\begin{equation}
 \sum_{n=N}^{\infty} E_n^{(i)}P_n^{(i)}
 =\hbar\omega_i r^N\left(N+\tfrac12+\frac{r}{1-r}\right).
 \label{eq:initial_moment_tail}
\end{equation}
At $\beta\hbar\omega_i=1$ and $N=40$ this is approximately
$1.75\times10^{-16}\hbar\omega_i$. Final deformed-energy moments are tested
separately in Sec.~\ref{sec:numerics}. We retain the notation $P_n^{(i,N)}$
when discussing truncation errors; finite sums below use the normalized finite
probabilities, with the superscript $N$ suppressed for readability.

During the first adiabatic stroke, the control vector is changed from $\lambda_i$ to
$\lambda_f$ while the working medium is isolated from the bath. The Hamiltonian,
Eq.~\eqref{eq:hamiltonian}, is a function of the number operator alone, and in the
realization $a_{p,q}=a_0\sqrt{[N]_{p,q}/N}$ the Fock states $|n\rangle$ are the same
vectors for all values of $(\omega,p,q)$. The labels in $|n;i\rangle$ and
$|n;f\rangle$ therefore identify the energies $E_n^{(i)}$ and $E_n^{(f)}$ of a common
eigenbasis. Consequently, $[H(\lambda),H(\lambda')]=0$ for all control vectors, the
evolution is diagonal in the Fock basis for any time dependence $\lambda(t)$, and the
populations are preserved exactly, independently of the driving speed. We use the
term ``adiabatic'' in the thermodynamic sense: the stroke exchanges no heat, and the
energy change of the working medium equals the work done on it. The stroke need not
be slow, and it produces neither coherences nor quantum friction. In a realization in
which the eigenbasis changes during the stroke, for example a trap whose frequency
is varied at fixed physical position, the Hamiltonians at different times would not
commute, and population preservation would require the quantum adiabatic condition
$\hbar\,|\langle m(t)|\partial_t n(t)\rangle|\ll|E_m(t)-E_n(t)|$, or a shortcut to
adiabaticity. This condition becomes more restrictive for $p,q<1$, where the level
spacings shrink at large $n$. The state after the first stroke is
\begin{equation}
 \rho_1=\sum_{n=0}^{N-1}P_n^{(i)}|n;f\rangle\langle n;f|.
 \label{eq:rho_after_first_adiabat}
\end{equation}
With the convention that positive work is work done on the working medium, the work in the first adiabatic stroke is
\begin{equation}
 W_I=\sum_{n=0}^{N-1}\left(E_n^{(f)}-E_n^{(i)}\right)P_n^{(i)}.
 \label{eq:work_I}
\end{equation}

For the finite-level model let $\Pi_N=\sum_{n=0}^{N-1}|n\rangle\langle n|$.
We specify $H_N=\sum_{n=0}^{N-1}E_n|n\rangle\langle n|$ and
$x_N=\Pi_N\hat{x}_{p,q}\Pi_N$, with no coupling across the upper boundary.
This projection preserves the prescribed spectral energies; it is not an exact
finite-dimensional representation of Eq.~\eqref{eq:pq_algebra}. If
$a_N=\Pi_N a_{p,q}\Pi_N$, the missing upper transition gives
\begin{equation}
 H_N=\frac{\hbar\omega}{2}(a_N^\dagger a_N+a_Na_N^\dagger)
 +\frac{\hbar\omega}{2}[N]_{p,q}|N-1\rangle\langle N-1|.
 \label{eq:finite_boundary}
\end{equation}
Thus Eq.~\eqref{eq:quadrature_hamiltonian} cannot be applied to the projected
quadratures without this boundary correction. All finite-level calculations use
$H_N$ and $x_N$ as specified here. This is a complete effective matrix model,
not a demonstrated microscopic realization.

\subsection{Gaussian measurement stroke}
\label{subsec:measurement_stroke}

A continuous set of Hermitian Kraus operators represents the measurement stroke,
\begin{equation}
 M_{\alpha}=\left(2\pi\sigma^2\right)^{-1/4}
 \exp\left[-\frac{\left(\hat{x}_{p,q}^{(f)}-\alpha\right)^2}{4\sigma^2}\right],
 \label{eq:measurement_operator}
\end{equation}
where $\hat{x}_{p,q}^{(f)}$ is evaluated with the final Hamiltonian parameters and $\sigma$ is the measurement resolution. The outcome $\alpha$, the quadrature $\hat{x}_{p,q}^{(f)}$, and the resolution $\sigma$ all have dimensions of length. In the numerical calculations, we use units with $\hbar=m=\omega_i=1$, so that lengths are measured in units of $\sqrt{\hbar/m\omega_i}$. The operators satisfy the completeness relation
\begin{equation}
 \int_{-\infty}^{\infty}M_{\alpha}^{\dagger}M_{\alpha}\,\dd\alpha=\Id.
 \label{eq:measurement_completeness}
\end{equation}
Because $[H_f, M_{\alpha}]\neq 0$ in general, the nonselective measurement changes the system energy. The postmeasurement state is
\begin{equation}
 \rho_2=\int_{-\infty}^{\infty}M_{\alpha}\rho_1M_{\alpha}\,\dd\alpha .
 \label{eq:postmeasurement_state}
\end{equation}

The natural oscillator length changes with $\omega_f$, so the same physical resolution corresponds to different effective measurement strengths at different final frequencies. The dimensionless resolution is
\begin{equation}
 \sigmat=\frac{\sigma}{x_0^{(f)}},
 \qquad
 x_0^{(f)}=\sqrt{\frac{\hbar}{m\omega_f}}.
 \label{eq:dimensionless_sigma}
\end{equation}
Throughout this work we hold the physical resolution $\sigma$ fixed, as for a measurement apparatus of given precision, and quote $\sigmat$ only for reference. With this convention, the undeformed measurement heat, Eq.~\eqref{eq:app_harmonic_measurement_benchmark}, is independent of $\omega_f$, which provides an exact check of all calculations involving a change of frequency (Appendix~\ref{app:scans}).

The transition matrix generated by the nonselective measurement is
\begin{equation}
 T_{nm}=\int_{-\infty}^{\infty}
 \left|\langle n;f|M_{\alpha}|m;f\rangle\right|^2\,\dd\alpha .
 \label{eq:transition_matrix}
\end{equation}
The postmeasurement populations in the $H_f$ basis are
\begin{equation}
 P_n^{(M)}=\langle n;f|\rho_2|n;f\rangle
 =\sum_{m=0}^{N-1}T_{nm}P_m^{(i)}.
 \label{eq:postmeasurement_populations}
\end{equation}
Completeness implies $\sum_n T_{nm}=1$ for each initial state $m$. For the Hermitian Gaussian measurement model, the channel also satisfies $T_{nm}=T_{mn}$; with the exact evaluation of $T_{nm}$ in Sec.~\ref{sec:numerics}, both properties hold identically for every $N$.
The measurement-induced heat is defined as the change in the oscillator energy during the nonselective measurement,
\begin{align}
 \QM
 &=\Tr\left[H_f(\rho_2-\rho_1)\right] \nonumber\\
 &=\sum_{n=0}^{N-1}E_n^{(f)}\left(P_n^{(M)}-P_n^{(i)}\right) \nonumber\\
 &=\sum_{n,m=0}^{N-1}\left(E_n^{(f)}-E_m^{(f)}\right)T_{nm}P_m^{(i)}.
 \label{eq:QM_general}
\end{align}
When $T_{nm}=T_{mn}$, this expression can be symmetrized as
\begin{equation}
 \QM=\frac{1}{2}\sum_{n,m=0}^{N-1}
 \left(E_n^{(f)}-E_m^{(f)}\right)T_{nm}
 \left(P_m^{(i)}-P_n^{(i)}\right).
 \label{eq:QM_symmetrized}
\end{equation}
The non-negativity of the transition probabilities alone does not prove $\QM>0$. A sufficient condition is that the state entering the measurement stroke be passive with respect to $H_f$ and that the nonselective measurement channel be unital \cite{Pusz1978CMP,Lenard1978JSP}. In the final energy basis, passivity requires
\begin{equation}
 E_n^{(f)}<E_m^{(f)} \quad \Rightarrow \quad P_n^{(i)}\geq P_m^{(i)}.
 \label{eq:passivity_condition}
\end{equation}
When the spectrum is ordered and $T_{nm}=T_{mn}$, Eq.~\eqref{eq:QM_symmetrized} becomes a sum of nonnegative pair contributions. At parameter points where these conditions do not hold, $\QM$ is evaluated directly, and engine operation is decided by the explicit conditions below. An explicit finite-dimensional proof of this passivity/unitality guarantee is provided in Appendix \ref{app:passivity_unitality}.

The same heat can be written in Heisenberg form as
\begin{equation}
 \QM=\sum_{m=0}^{N-1}P_m^{(i)}
 \langle m;f|\left(H_M-H_f\right)|m;f\rangle,
 \label{eq:heisenberg_QM}
\end{equation}
where
\begin{equation}
 H_M=\int_{-\infty}^{\infty}M_{\alpha}H_f M_{\alpha}\,\dd\alpha.
 \label{eq:HM}
\end{equation}
Insertion of the final energy resolution of the identity in Eq.~\eqref{eq:heisenberg_QM} recovers Eq.~\eqref{eq:QM_general}.

\subsection{Second adiabatic stroke, thermalization, and efficiency}
\label{subsec:second_stroke}

After the measurement, the control parameters are changed adiabatically from $\lambda_f$ back to $\lambda_i$. As in the first stroke, the Hamiltonians commute, so the populations $P_n^{(M)}$ are preserved exactly during this reverse adiabatic stroke, while the eigenvalues return to $E_n^{(i)}$. The associated work is
\begin{equation}
 W_{II}=\sum_{n=0}^{N-1}\left(E_n^{(i)}-E_n^{(f)}\right)P_n^{(M)}.
 \label{eq:work_II}
\end{equation}
The final thermalization stroke returns the system to $\rho_0$. The heat exchanged with the bath, positive when absorbed by the system, is
\begin{equation}
 \QT=\sum_{n=0}^{N-1}E_n^{(i)}\left(P_n^{(i)}-P_n^{(M)}\right).
 \label{eq:thermalization_heat}
\end{equation}
For engine operation, $\QT<0$ indicates heat rejection to the bath.

The first-law balance over a complete cycle is
\begin{equation}
 W_I+\QM+W_{II}+\QT=0.
 \label{eq:first_law}
\end{equation}
The net work done on the system is
\begin{equation}
 \Wcyc=W_I+W_{II},
 \label{eq:Wcyc}
\end{equation}
Moreover, the useful output is $\Wout=-\Wcyc$. The working-medium efficiency is
\begin{equation}
 \eta=\frac{\Wout}{\QM}=\frac{-\Wcyc}{\QM},
 \qquad
 \QM>0,
 \qquad
 \Wcyc<0.
 \label{eq:efficiency}
\end{equation}
The physically admissible engine regime is selected by
\begin{equation}
 \QM>0,
 \qquad
 \Wcyc<0,
 \qquad
 \QT<0,
 \qquad
 0<\eta<1.
 \label{eq:engine_conditions}
\end{equation}
For the undeformed oscillator, the cycle reproduces the quasistatic Otto efficiency
exactly. Writing $\delta P_n=P_n^{(M)}-P_n^{(i)}$, Eqs.~\eqref{eq:QM_general} and (\ref{eq:work_pair_appendix}) give
$\QM=\sum_n E_n^{(f)}\delta P_n$ and
$\Wcyc=-\sum_n\big(E_n^{(f)}-E_n^{(i)}\big)\delta P_n$. If the spectrum is scaled
uniformly, $E_n^{(i)}=(\omega_i/\omega_f)E_n^{(f)}$ for all $n$, then
$\Wcyc=-(1-\omega_i/\omega_f)\QM$ and
\begin{equation}
 \eta_{\rm Otto}=1-\frac{\omega_i}{\omega_f},
 \label{eq:eta_otto}
\end{equation}
independently of the measurement strength, the temperature, and the form of the
population change $\delta P_n$~\cite{Yi2017PRE}. The Otto efficiency therefore does not
depend on whether the energy is supplied by a hot bath or by a measurement channel;
it depends only on the uniform scaling of the spectrum. The deformation breaks this
scaling. For a general spectrum, the same relations give the exact expression
\begin{equation}
 \eta=1-\frac{\sum_n E_n^{(i)}\,\delta P_n}{\sum_n E_n^{(f)}\,\delta P_n}
     =1+\frac{\QT}{\QM}.
 \label{eq:eta_general}
\end{equation}
The exact condition for $\eta>1-r$, where $r=\omega_i/\omega_f$ and $\QM>0$,
is $\sum_n(E_n^{(i)}-rE_n^{(f)})\delta P_n<0$. It is invariant under independent
constant shifts of either Hamiltonian. Individual ratios $E_n^{(f)}/E_n^{(i)}$
are not such an invariant criterion. For the symmetric transition matrix one may
instead use pairwise gaps: $\QM=\sum_{n>m}(E_n^{(f)}-E_m^{(f)})T_{nm}(P_m-P_n)$,
with an analogous sum for the initial-energy change. It is this gap-weighted
redistribution, rather than an absolute energy ratio, that controls the efficiency.

\section{Spectral validity, numerical implementation, and convergence}
\label{sec:numerics}

The cycle uses a Gibbs state only once, at the start, and only of the initial
Hamiltonian $H_i$. In this work $(p_i,q_i)=(1,1)$, so $H_i$ is the ordinary harmonic
oscillator and $Z_i$ is always finite. The final Hamiltonian $H_f$ is never
thermalized. Its spectrum enters the energetics only through the eigenvalues
$E_n^{(f)}$, weighted by the populations $P_n^{(i)}$ and $P_n^{(M)}$ in
Eqs.~\eqref{eq:work_I}-- \eqref{eq:thermalization_heat}. No partition function of $H_f$ is required. The infinite-oscillator
quantities are well defined whenever these weighted sums converge, and the measurement
itself is uniquely defined, a condition discussed below.

The structure of the final spectrum matters differently. For $p=q=s<1$, the
diagonal limit $[n]_{s,s}=ns^{n-1}$ first increases and then decreases with $n$, so
the final spectrum is bounded and not monotonic (Appendix~\ref{app:checks}). For a finite retained basis, this loss of ordering occurs only when the turning
point lies below its upper boundary; it does not follow from $\max(p_f,q_f)<1$
alone for parameters arbitrarily close to unity. At such points the
ordering condition
\begin{equation}
 E_{n+1}^{(f)}>E_n^{(f)},\qquad n=0,\ldots,N-2,
 \label{eq:spectral_order_check}
\end{equation}
fails, and the passivity argument of Appendix~\ref{app:passivity_unitality}, which guarantees
$\QM\ge0$, does not apply. We do not exclude these points. Instead, $\QM$ is
evaluated directly from Eq.~\eqref{eq:QM_general}, and engine operation is decided by the
explicit conditions in Eq.~\eqref{eq:engine_conditions}. Where
Eq.~\eqref{eq:spectral_order_check} holds, $\QM\ge0$ is guaranteed analytically.

Because the Kraus operators in Eq.~\eqref{eq:measurement_operator} are functions of
the measured quadrature, $M_\alpha=g(\hat{x}^{(f)}_{p,q}-\alpha)$ with
$g(x)=(2\pi\sigma^2)^{-1/4}e^{-x^2/4\sigma^2}$, the integral over $\alpha$ in
Eq.~\eqref{eq:transition_matrix} can be performed exactly. Let
$\hat{x}^{(f)}_{p,q}|x_k\rangle=x_k|x_k\rangle$, $k=1,\dots,N$, denote the eigenbasis
of the quadrature in the retained $N$-level space, and define
\begin{equation}
 u_k^{(nm)}=\langle n;f|x_k\rangle\langle x_k|m;f\rangle,
 \qquad
 K_{kl}=e^{-(x_k-x_l)^2/8\sigma^2}.
 \label{eq:u_K_def}
\end{equation}
Using $\int_{-\infty}^{\infty}g(x_k-\alpha)\,g(x_l-\alpha)\,d\alpha=K_{kl}$, the
transition matrix becomes
\begin{equation}
 T_{nm}=\sum_{k,l=1}^{N}u_k^{(nm)}\,K_{kl}\,u_l^{(nm)*}.
 \label{eq:Texact}
\end{equation}
Equation~\eqref{eq:Texact} replaces numerical quadrature and contains no
discretization error in $\alpha$. It is manifestly symmetric and doubly stochastic
for every $N$, so the normalization and symmetry errors
\begin{equation}
 \epsilon_{\mathrm{norm}}=\max_m\Big|\sum_n T_{nm}-1\Big|,
 \qquad
 \epsilon_{\mathrm{sym}}=\max_{n,m}\big|T_{nm}-T_{mn}\big|
 \label{eq:norm_sym_error}
\end{equation}
vanish to machine precision. The first-law error
\begin{equation}
 \epsilon_{\mathrm{FL}}=\left|W_I+\QM+W_{II}+\QT\right|
 \label{eq:first_law_error}
\end{equation}
is monitored as an independent check of the sign convention.
Equivalently, the nonselective measurement acts on the state as
\begin{equation}
 \Phi(\rho)=\exp\!\left(-\frac{\mathcal{D}_x^2}{8\sigma^2}\right)\rho,
 \qquad
 \mathcal{D}_x(A)\equiv\big[\hat{x}^{(f)}_{p,q},A\big],
 \label{eq:channel_superop}
\end{equation}
which, in the specified finite-dimensional model, is the operator form of the dephasing kernel in Eq.~\eqref{eq:app_gaussian_channel_kernel}. Since
$\mathcal{D}_x^2$ is self-adjoint with respect to the Hilbert--Schmidt inner product,
the measurement heat can be written exactly as
\begin{equation}
 \QM=\sum_{k=1}^{\infty}\frac{(-1)^k}{k!\,(8\sigma^2)^k}\,
 \mathrm{Tr}\!\left[\rho_1\,\mathcal{D}_x^{2k}(H_f)\right].
 \label{eq:QM_series}
\end{equation}
Formally evaluating the $k=1$ term before truncation gives the weak-measurement result
\begin{equation}
 \QM=\frac{\hbar}{8m\omega_f\sigma^2}\sum_{n=0}^{\infty}P_n^{(i)}B_n
 +O\!\left(\sigma^{-4}\right),
 \label{eq:QM_weak}
\end{equation}
where
\begin{equation}
\begin{aligned}
B_n &= [n+1]_{p_f,q_f}\,\Delta_n^{(f)} \\
    &\quad - [n]_{p_f,q_f}\,\Delta_{n-1}^{(f)}, \\
\Delta_n^{(f)} &= E_{n+1}^{(f)}-E_n^{(f)}.
\end{aligned}
\label{eq:Bn_def}
\end{equation}
with $[0]_{p,q}\Delta_{-1}^{(f)}\equiv0$. For the finite matrix model, the sum ends at
$N-1$ and its upper-boundary term is instead
$B_{N-1}^{(N)}=-[N-1]_{p_f,q_f}\Delta_{N-2}^{(f)}$; the upward transition is absent.
The series in Eq.~\eqref{eq:QM_series} converges for each fixed finite $N$.
For an unbounded infinite operator, interchanging its series with an energy trace
requires additional domain and summability assumptions. For the undeformed oscillator,
$\Delta_n^{(f)}=\hbar\omega_f$ and $B_n=\hbar\omega_f$ for every $n$, so
Eq.~\eqref{eq:QM_weak} reduces to the exact, state-independent result
$\hbar^2/8m\sigma^2$ of Eq.~\eqref{eq:app_harmonic_measurement_benchmark}. For the deformed oscillator, each level contributes
a different amount of measurement heat, weighted by the product of a deformed number
and a deformed level spacing. At the parameters of
Table~\ref{tab:numerical_parameters}, Eq.~\eqref{eq:QM_weak} agrees with the
numerical $\QM$ of the $N=40$ model to within $1.1\%$ throughout the retained domain.

Truncating the Hilbert space at dimension $N$ is a controlled approximation to the
infinite oscillator only if the sums in Eqs.~\eqref{eq:work_I}-- \eqref{eq:thermalization_heat} converge.
Two mechanisms limit this convergence. The first is the thermal tail of the initial
state. For $s=\max(p_f,q_f)>1$ the deformed energies grow as $E_n^{(f)}\propto s^n$,
while $P_n^{(i)}\propto e^{-\beta\hbar\omega_i n}$. The first adiabatic work
$W_I=\sum_n(E_n^{(f)}-E_n^{(i)})P_n^{(i)}$ therefore converges only if
\begin{equation}
 \max(p_f,q_f)<e^{\beta\hbar\omega_i}.
 \label{eq:thermal_criterion}
\end{equation}
The second mechanism is specific to the measurement. In Eq.~\eqref{eq:Bn_def} both
$[n]_{p,q}$ and $\Delta_n^{(f)}$ grow as $s^n$, so $B_n$ grows as $s^{2n}$. The
leading measurement heat therefore converges only if
\begin{equation}
 \max(p_f,q_f)<e^{\beta\hbar\omega_i/2},
 \label{eq:measurement_criterion}
\end{equation}
which is more restrictive than Eq.~\eqref{eq:thermal_criterion}. These strict
inequalities also hold on the diagonal $p_f=q_f$, where polynomial factors in $n$
multiply the exponential growth. Equation~\eqref{eq:measurement_criterion} concerns
only the leading weak-measurement coefficient. It neither proves convergence nor
proves divergence of the exact finite-strength measurement heat. Higher-order
coefficients can require stronger moment bounds; factorial suppression alone does
not justify an interchange with an infinite energy sum.

There is a more fundamental reason why truncation cannot always reproduce an infinite
oscillator, which concerns the measured operator itself. In the Fock basis,
$\hat{x}^{(f)}_{p,q}$ is a Jacobi (tridiagonal) matrix with vanishing diagonal and
off-diagonal elements $b_n=\sqrt{\hbar/2m\omega_f}\,\sqrt{[n+1]_{p,q}}$. The deformed
numbers satisfy $[n]_{p,q}^2-[n-1]_{p,q}[n+1]_{p,q}=(pq)^{n-1}>0$, so that
$b_{n-1}b_{n+1}\le b_n^2$ for all $n$. By Carleman's criterion, the operator is
essentially self-adjoint if $\sum_n b_n^{-1}$ diverges, and by Berezanskii's theorem for a bounded Jacobi diagonal
it is not essentially self-adjoint if the sum converges and $b_{n-1}b_{n+1}\le b_n^2$~\cite{Reiffenstein2025}.
For $\max(p,q)\le1$ one has $[n]_{p,q}\le n$, the sum diverges, and
$\hat{x}^{(f)}_{p,q}$ defines a unique self-adjoint quadrature. For $\max(p,q)>1$ one
has $[n]_{p,q}\ge\max(p,q)^{n-1}$, the sum converges, and the quadrature admits
infinitely many self-adjoint extensions (Appendix~\ref{subsec:finite_truncation}).
The Gaussian measurement of Eq.~\eqref{eq:measurement_operator}, and hence $\QM$,
$W_{II}$, and $\eta$, is then not uniquely defined for the infinite oscillator. Each
truncation selects a particular finite-dimensional quadrature, and quantities that
weight highly excited levels, such as $\QM$, need not have a limit as $N\to\infty$.
For $\max(p_f,q_f)>1$ we therefore choose an explicitly specified effective
$N$-level system, the truncated deformed oscillator, and choose parameters for which
its properties are insensitive to the value of $N$ over a wide range. The first
adiabatic work $W_I$ is independent of the quadrature-extension choice, but its
energy-moment convergence must still be checked.

At the original $\beta\hbar\omega_i=0.2$, an $N=10$ basis omits
$e^{-2}=0.135335$ of the initial probability. The leading measurement coefficient
already fails its moment test for $\max(p_f,q_f)\ge e^{0.1}$, but this alone does
not settle the exact-channel limit. We withdraw the former infinite-oscillator
interpretation of large-deformation results. The revised calculations use
$\beta\hbar\omega_i=1$, $\sigma=4$, $N=40$, and $0.7\le p_f,q_f\le1.2$.
For $\max(p_f,q_f)>1$, increasing $N$ compares different effective systems rather
than controlling an approximation to a specified infinite oscillator. The range is
chosen for modest dimension sensitivity at the reported parameters, not as a proved
convergence domain. Larger deformations, such as $(1.3,1.3)$, are more sensitive.

The parameters used in all figures are listed in
Table~\ref{tab:numerical_parameters}. All quantities are dimensionless, with
$\hbar=k_{\mathrm B}=m=1$. The measurement resolution $\sigma$ is held fixed in
physical units;  the corresponding dimensionless resolution of Eq.~\eqref{eq:dimensionless_sigma}, $\tilde\sigma=\sigma\sqrt{m\omega_f/\hbar}$, ranges from $4$ to $4\sqrt3$ across the frequency scan of Appendix~\ref{app:scans}.

\begin{table}[tbp]
\caption{Numerical parameters used in the simulations.}
\label{tab:numerical_parameters}
\begin{ruledtabular}
\begin{tabular}{p{0.33\columnwidth}p{0.55\columnwidth}}
\raggedright Quantity & \raggedright Value used in figures \tabularnewline
\hline
\raggedright Energy units & \raggedright $\hbar=1$, $k_{\mathrm B}=1$ \tabularnewline
\raggedright Mass & \raggedright $m=1$ \tabularnewline
\raggedright Initial frequency & \raggedright $\omega_i=1$ \tabularnewline
\raggedright Final frequency & \raggedright $\omega_f/\omega_i=2$ \tabularnewline
\raggedright Frequency scan & \raggedright $1\le\omega_f/\omega_i\le3$ \tabularnewline
\raggedright Initial deformation & \raggedright $(p_i,q_i)=(1,1)$ \tabularnewline
\raggedright Final deformation range & \raggedright $0.7\le p_f,q_f\le1.2$ \tabularnewline
\raggedright Inverse temperature & \raggedright $\beta\hbar\omega_i=1$ \tabularnewline
\raggedright Measurement resolution & \raggedright $\sigma=4$ (physical; $\sigmat=4\sqrt{2}$ at $\omega_f=2\omega_i$) \tabularnewline
\raggedright Hilbert-space dimension & \raggedright $N=40$ \tabularnewline
\raggedright Transition matrix & \raggedright exact, Eq.~\eqref{eq:Texact} \tabularnewline
\raggedright Deformation grid & \raggedright $39	\times 39=1521$ points for contour plots \tabularnewline
\end{tabular}
\end{ruledtabular}
\end{table}

We test the numerical results in two ways. First, for the undeformed final oscillator, all
quantities are known in closed form:
\begin{equation}
\begin{aligned}
 \QM^{\rm HO}&=\frac{\hbar^2}{8m\sigma^2},\\
 W_I^{\rm HO}&=\hbar(\omega_f-\omega_i)\left(\bar n+\tfrac12\right),\\
 \eta^{\rm HO}&=1-\frac{\omega_i}{\omega_f},\\
 W_{II}^{\rm HO}&=-W_I^{\rm HO}-\left(1-\frac{\omega_i}{\omega_f}\right)\QM^{\rm HO}.
\end{aligned}
 \label{eq:HO_exact}
\end{equation}
with $\bar n=(e^{\beta\hbar\omega_i}-1)^{-1}$. At the parameters of
Table~\ref{tab:numerical_parameters}, the numerical method reproduces all four
quantities in every displayed digit (first row of Table~\ref{tab:convergence}).
Second, for the deformed oscillator we compare the results at $N=40$ and $N=60$,
computed with 50-digit arithmetic, through the absolute and relative changes
\begin{equation}
 \Delta_N X=\left|X^{(60)}-X^{(40)}\right|,
 \qquad
 r_N(X)=\frac{\Delta_N X}{\left|X^{(60)}\right|},
 \label{eq:conv_measures}
\end{equation}
for $X\in\{\QM,W_I,W_{II},\eta\}$. Table~\ref{tab:convergence} lists these errors at
representative points of the domain, including the corner $p_f=q_f=1.2$, where the
working-medium efficiency is maximal.

\begin{table*}[tp]
\caption{Dimension sensitivity at the parameters of Table~\ref{tab:numerical_parameters}.
The entries use independently evaluated 50-digit arithmetic. For each point and
observable $X$, the table gives $X^{(40)}$, $|X^{(60)}-X^{(40)}|$, and that absolute
change divided by $|X^{(60)}|$. These are changes between specified finite models,
not certified errors relative to an infinite oscillator. Efficiency is omitted
outside the engine domain.}
\label{tab:convergence}
\begin{ruledtabular}
\begin{tabular}{ccrrr}
$(p_f,q_f)$ & $X$ & $X^{(40)}$ & absolute change & relative change \\
\hline
$(1,1)$ & $Q_{\rm M}$ & $0.007812500$ & $2.60\times10^{-18}$ & $3.33\times10^{-16}$ \\
$(1,1)$ & $W_I$ & $1.081976707$ & $1.70\times10^{-16}$ & $1.57\times10^{-16}$ \\
$(1,1)$ & $W_{II}$ & $-1.085882957$ & $1.71\times10^{-16}$ & $1.58\times10^{-16}$ \\
$(1,1)$ & $W_{\rm cyc}$ & $-0.003906250$ & $1.30\times10^{-18}$ & $3.33\times10^{-16}$ \\
$(1,1)$ & $\eta$ & $0.500000000$ & $6.00\times10^{-50}$ & $1.20\times10^{-49}$ \\
$(0.7,0.7)$ & $Q_{\rm M}$ & $0.002499775$ & $1.06\times10^{-20}$ & $4.25\times10^{-18}$ \\
$(0.7,0.7)$ & $W_I$ & $0.486481971$ & $1.77\times10^{-16}$ & $3.63\times10^{-16}$ \\
$(0.7,0.7)$ & $W_{II}$ & $-0.486152912$ & $1.77\times10^{-16}$ & $3.63\times10^{-16}$ \\
$(0.7,0.7)$ & $W_{\rm cyc}$ & $0.000329058$ & $1.40\times10^{-21}$ & $4.25\times10^{-18}$ \\
$(0.7,0.7)$ & $\eta$ & --- & --- & --- \\
$(0.7,1.2)$ & $Q_{\rm M}$ & $0.007038985$ & $2.05\times10^{-11}$ & $2.92\times10^{-9}$ \\
$(0.7,1.2)$ & $W_I$ & $1.003006192$ & $3.09\times10^{-14}$ & $3.08\times10^{-14}$ \\
$(0.7,1.2)$ & $W_{II}$ & $-1.006281697$ & $2.06\times10^{-11}$ & $2.04\times10^{-11}$ \\
$(0.7,1.2)$ & $W_{\rm cyc}$ & $-0.003275505$ & $2.05\times10^{-11}$ & $6.27\times10^{-9}$ \\
$(0.7,1.2)$ & $\eta$ & $0.465337671$ & $1.56\times10^{-9}$ & $3.35\times10^{-9}$ \\
$(1,1.2)$ & $Q_{\rm M}$ & $0.013110837$ & $2.58\times10^{-9}$ & $1.97\times10^{-7}$ \\
$(1,1.2)$ & $W_I$ & $1.367029636$ & $7.75\times10^{-14}$ & $5.67\times10^{-14}$ \\
$(1,1.2)$ & $W_{II}$ & $-1.375717336$ & $2.58\times10^{-9}$ & $1.88\times10^{-9}$ \\
$(1,1.2)$ & $W_{\rm cyc}$ & $-0.008687699$ & $2.58\times10^{-9}$ & $2.97\times10^{-7}$ \\
$(1,1.2)$ & $\eta$ & $0.662634995$ & $6.64\times10^{-8}$ & $1.00\times10^{-7}$ \\
$(1.2,1.2)$ & $Q_{\rm M}$ & $0.023018401$ & $2.03\times10^{-5}$ & $8.82\times10^{-4}$ \\
$(1.2,1.2)$ & $W_I$ & $1.689631975$ & $5.35\times10^{-13}$ & $3.17\times10^{-13}$ \\
$(1.2,1.2)$ & $W_{II}$ & $-1.707640633$ & $2.03\times10^{-5}$ & $1.19\times10^{-5}$ \\
$(1.2,1.2)$ & $W_{\rm cyc}$ & $-0.018008658$ & $2.03\times10^{-5}$ & $1.13\times10^{-3}$ \\
$(1.2,1.2)$ & $\eta$ & $0.782359188$ & $1.92\times10^{-4}$ & $2.45\times10^{-4}$ \\
\end{tabular}
\end{ruledtabular}
\end{table*}

The $N=40$ to $60$ first-stroke work changes are below $10^{-12}$ at the
listed points. At $(1.2,1.2)$ the relative changes are $8.82\times10^{-4}$ in
$\QM$, $1.19\times10^{-5}$ in $W_{II}$, $1.13\times10^{-3}$ in net work, and
$2.45\times10^{-4}$ in $\eta$. A separate double-precision check covers every point
of the regenerated $39\times39$ map: the largest absolute changes are
$2.04\times10^{-5}$ in $\QM$, $5.4\times10^{-13}$ in $W_I$,
$2.04\times10^{-5}$ in $W_{II}$ and $\Wcyc$, and $1.92\times10^{-4}$ in $\eta$.
Relative errors near a zero of net work or efficiency are not useful, so the
absolute differences should be used there. These tests do not certify a continuum
of parameter points between grid nodes.

The sensitivity is also nonuniform with dimension: at the corner $\QM$ changes
by approximately $0.67\%$ between $N=40$ and $80$. The $40$ to $60$ difference is
therefore not a bound on an infinite-dimensional error. At $N=10$, $\eta=0.7747$
is close to its $N=40$ value, but $\QM$ differs by about $3.8\%$ and $W_I$ by
about $0.36\%$; efficiency stability must not be generalized to all observables.
All subsequent maps describe the explicitly specified $N=40$ effective model.
For $\max(p_f,q_f)\le1$, increasing the cutoff can instead approximate the unique
infinite quadrature, subject to convergence of the relevant energy traces.

\section{Results}
\label{sec:results}

\subsection{Dependence on the Hilbert-space dimension}

\begin{figure}[tbp]
 \centering
\includegraphics[width=\columnwidth]{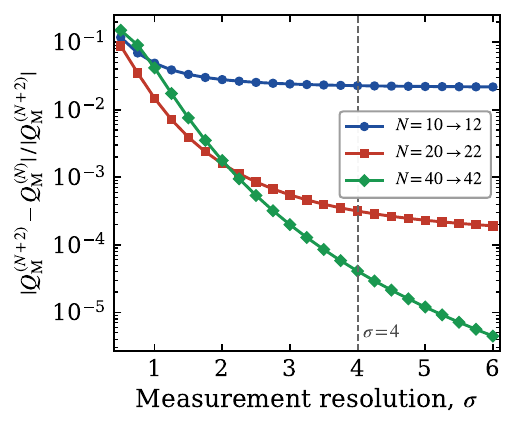}
 \caption{Relative change with finite dimension of the measurement-induced heat,
$|Q_{\rm M}^{(N+2)}-Q_{\rm M}^{(N)}|/|Q_{\rm M}^{(N+2)}|$, as a function of the
measurement resolution $\sigma$ for three successive truncations, at $\beta\hbar\omega_i=1$, $\omega_f/\omega_i=2$, $p_i=q_i=1$, and $p_f=q_f=1.2$.
The transition matrix is evaluated exactly, Eq.~\eqref{eq:Texact}. For weak
measurements, the error decreases rapidly with $N$; at the value $\sigma=4$ used
in this work (dashed line), it is $4\times10^{-5}$ for $N=40$. For strong
measurements ($\sigma\lesssim2$), the error does not decrease with $N$, showing greater sensitivity to high-lying retained levels.
These are finite-model dimension comparisons, not infinite-limit error bounds.}
 \label{fig:convergence_QM}
\end{figure}

Figure~\ref{fig:convergence_QM} shows the relative cutoff error of the
measurement-induced heat, $|\QM^{(N+2)}-\QM^{(N)}|/|\QM^{(N+2)}|$, as a function of
the measurement resolution $\sigma$, at the corner $p_f=q_f=1.2$ of the deformation
domain, where the dependence on $N$ is strongest. Using the relative rather than the
absolute error is essential here. Since $\QM$ itself decreases approximately as $\sigma^{-2}$,
Eq.~\eqref{eq:QM_weak}, absolute differences shrink with $\sigma$ even when the
truncation is not converged.

For weak measurements, the successive-dimension differences are small over the
illustrated range. At $\sigma=4$, the $N=40$ to $42$ relative change is about
$4\times10^{-5}$. Stronger measurement increases sensitivity to the retained
high-energy levels. These curves characterize the finite models; a decrease over
a finite set of cutoffs does not establish an infinite-oscillator limit. Table~\ref{tab:convergence}
gives absolute and relative changes for all cycle observables at representative points,
including the corner of maximum reported efficiency. Net work is particularly
sensitive because it is a small difference of larger stroke works.

\subsection{Measurement-induced heat}
\label{subsec:QM_results}

\begin{figure}[tbp]
 \centering
 \includegraphics[width=\columnwidth]{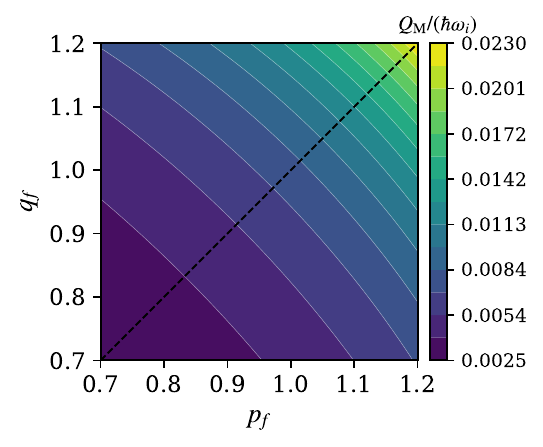}
 \caption{Measurement-induced heat $\QM$ in the $(p_f,q_f)$ plane at
$\beta\hbar\omega_i=1$, $\omega_f/\omega_i=2$, $\sigma=4$, and $N=40$. The dashed
line marks the diagonal $p_f=q_f$; the map is symmetric about it because
$[n]_{p,q}=[n]_{q,p}$. The undeformed oscillator, $p_f=q_f=1$, gives the exact value
$\QM^{\rm HO}=\hbar^2/8m\sigma^2=0.0078$. The heat increases toward the upper-right
corner, where both deformation parameters are large.}
 \label{fig:QM_contour}
\end{figure}

Figure~\ref{fig:QM_contour} shows the measurement-induced heat over the retained
deformation domain. $\QM$ increases monotonically with both $p_f$ and $q_f$, from
about $0.32\,\QM^{\rm HO}$ at $(0.7,0.7)$ to about $2.9\,\QM^{\rm HO}$ at
$(1.2,1.2)$. This follows from the weak-measurement expression,
Eq.~\eqref{eq:QM_weak}: each level contributes in proportion to
$B_n=[n+1]_{p_f,q_f}\Delta_n^{(f)}-[n]_{p_f,q_f}\Delta_{n-1}^{(f)}$. For the
undeformed oscillator, $B_n=\hbar\omega_f$ for every level, so the heat is independent
of the state. Deformation with $p_f,q_f>1$ widens both the deformed numbers and the
level spacings, and the measurement then injects more energy into each occupied
level. For $p_f,q_f<1$ the spacings shrink and $\QM$ falls below the undeformed
value. The heat remains positive throughout the domain, including the region
$\max(p_f,q_f)<1$, where the final spectrum is not ordered and the passivity
guarantee of Appendix~\ref{app:passivity_unitality} does not apply.

The shape of the contours shows that the two deformation parameters do not act only
through their sum. Close to $(1,1)$ the contours are nearly perpendicular to the
diagonal, so $\QM$ depends mainly on $\delta=(p_f+q_f)/2-1$, as predicted by the
first-order expansion of Appendix~\ref{app:perturbative_expansion}. Away from this point, the contours bend toward
the origin. At fixed $p_f+q_f$, an asymmetric pair therefore produces more heat than
a symmetric one. For example, $(1.2,0.8)$ and $(1,1)$ both have $\delta=0$, yet
$\QM(1.2,0.8)=0.0085$ exceeds $\QM^{\rm HO}=0.0078$. This dependence on the
antisymmetric combination $\gamma=(p_f-q_f)/2$ appears only beyond first order and
is absent at fixed $\delta=0$ on the standard one-parameter slices considered here.

\subsection{Adiabatic work contributions}
\label{subsec:work_results}

\begin{figure}[tbp]
 \centering
 \includegraphics[width=\columnwidth]{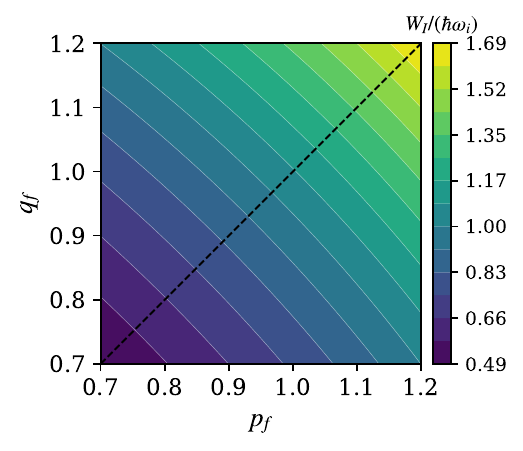}
 \caption{First adiabatic work $W_I$ in the $(p_f,q_f)$ plane at
$\beta\hbar\omega_i=1$, $\omega_f/\omega_i=2$, and $N=40$. Positive values denote
work done on the working medium during the compression-like stroke. $W_I$ is
independent of the measurement. The dashed line marks the diagonal $p_f=q_f$.}
\label{fig:WI_contour}
\end{figure}

\begin{figure}[tbp]
 \centering
 
\includegraphics[width=\columnwidth]{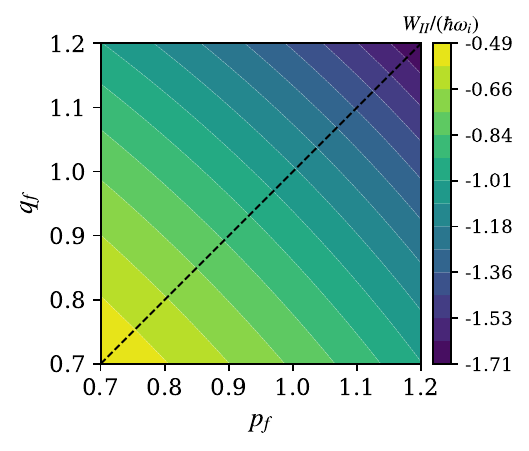}
 \caption{Second adiabatic work $W_{II}$ in the $(p_f,q_f)$ plane at
$\beta\hbar\omega_i=1$, $\omega_f/\omega_i=2$, $\sigma=4$, and $N=40$. Negative values
denote work extracted from the working medium during the return stroke; the yellow
region marks the largest extracted work. The dashed line marks the diagonal
$p_f=q_f$.}
\label{fig:WII_contour}
\end{figure}

Figures~\ref{fig:WI_contour} and \ref{fig:WII_contour} show the two adiabatic work
contributions over the retained domain. The first adiabatic work,
$W_I=\sum_n(E_n^{(f)}-E_n^{(i)})P_n^{(i)}$, depends only on the change of spectrum and
on the initial thermal populations, and is independent of the measurement. It
increases with both deformation parameters, from $W_I=0.486$ at $(0.7,0.7)$ to
$W_I=1.690$ at $(1.2,1.2)$, because $[n]_{p,q}$ is a polynomial in $p$ and $q$ with
positive coefficients, Eq.~\eqref{eq:pq_sum}, so every final level rises as $p_f$ or $q_f$
increases. $W_I$ remains positive even where $p_f,q_f<1$ lower the spectrum, because
the frequency compression $\omega_f=2\omega_i$ dominates there. Near the undeformed
point, the perturbative result of Eq.~\eqref{eq:pert_work_I} reproduces the numerical values well, to
within $2.4\%$ at $(1.0,1.2)$, and to within $6\%$ even at the corner $(1.2,1.2)$.

The second adiabatic work closely mirrors the first, ranging from $W_{II}=-0.486$ at
$(0.7,0.7)$ to $W_{II}=-1.708$ at $(1.2,1.2)$. This follows directly from the
definitions: $W_{II}=\Wcyc-W_I$. Since the measurement redistributes only a small
fraction of the population, $|\Wcyc|$ is about two orders of magnitude smaller than
$W_I$ throughout the domain. For example, at $(1.2,1.2)$ one has $\Wcyc=-0.018$,
compared with $W_I=1.690$. The return stroke therefore recovers almost all of the work
invested in the first stroke, and engine operation is determined by the small excess
extracted because the measurement has moved population into higher levels before the
return stroke. 

As for $\QM$, the contours of both work contributions are nearly perpendicular to the
diagonal close to $(1,1)$, where the first-order dependence on $\delta$ dominates, and
bend toward the origin further away. At fixed $\delta$, an asymmetric pair of
deformation parameters therefore gives a larger work contribution than a symmetric
one; for example, $W_I(1.2,0.8)=1.112$ exceeds $W_I(1,1)=1.082$.

\subsection{Net work and efficiency}
\label{subsec:efficiency_results}

\begin{figure}[tbp]
 \centering
 \includegraphics[width=\columnwidth]{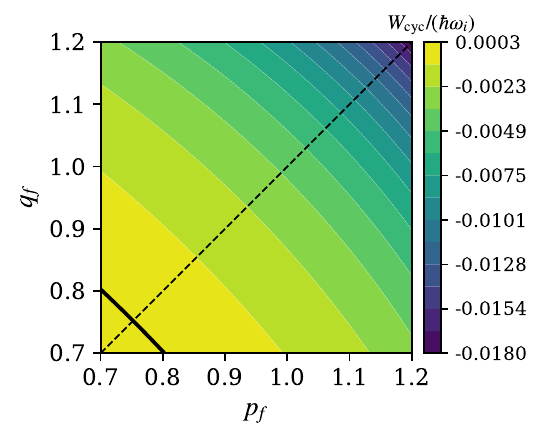}
 \caption{Net work $\Wcyc=W_I+W_{II}$ in the $(p_f,q_f)$ plane at
$\beta\hbar\omega_i=1$, $\omega_f/\omega_i=2$, $\sigma=4$, and $N=40$. Negative values
correspond to engine operation; the yellow region marks the largest work output. The
thick black line is the contour $\Wcyc=0$. Below it, in the lower-left corner, net
work is done on the system. The dashed line marks the diagonal $p_f=q_f$.}
\label{fig:Wcyc_contour}
\end{figure}

The net work, $\Wcyc=W_I+W_{II}$, is shown in Fig.~\ref{fig:Wcyc_contour}. It is
negative over almost the entire retained domain, so the cycle operates as an engine.
The only exception is a small region in the lower-left corner, bounded approximately
by $p_f+q_f\approx1.5$, where the deformation lowers the final spectrum so strongly
that net work is done on the system; at $(0.7,0.7)$, $\Wcyc=+3.3\times10^{-4}$. The
useful work output $W_{\rm out}=-\Wcyc$ increases with both deformation parameters
and reaches $0.0180$ at $(1.2,1.2)$, $4.6$ times the undeformed value
$W_{\rm out}^{\rm HO}=(1-\omega_i/\omega_f)\QM^{\rm HO}=0.0039$. As discussed above,
$|\Wcyc|$ is about two orders of magnitude smaller than the individual stroke works,
because it is the small excess that the return stroke extracts from the population
moved upward by the measurement.

\begin{figure}[tbp]
 \centering
 \includegraphics[width=\columnwidth]{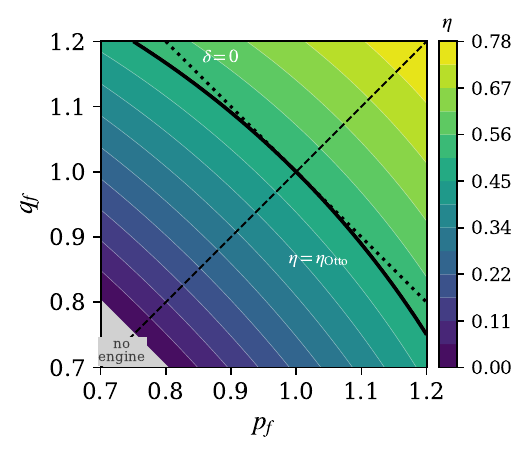}
 \caption{Working-medium efficiency $\eta=-\Wcyc/\QM$ in the $(p_f,q_f)$ plane at the
parameters of Table~\ref{tab:numerical_parameters}. The grey region in the lower-left
corner violates the engine conditions of Eq.~\eqref{eq:engine_conditions} and is
excluded. The thick black line is the contour $\eta=\eta_{\rm Otto}=1-\omega_i/\omega_f=0.5$;
above it the deformed engine is more efficient than the undeformed oscillator. The
dotted line marks $\delta=(p_f+q_f)/2-1=0$, and the dashed line the diagonal
$p_f=q_f$. The maximum, $\eta=0.782$, occurs at $p_f=q_f=1.2$.}
\label{fig:eta_contour}
\end{figure}

Figure~\ref{fig:eta_contour} shows the working-medium efficiency. Because $\QM>0$
throughout the domain, the contour $\eta=0$ coincides with $\Wcyc=0$, and the region
violating the engine conditions of Eq.~\eqref{eq:engine_conditions} is the same
lower-left corner. The undeformed oscillator gives exactly the Otto efficiency,
$\eta_{\rm Otto}=1-\omega_i/\omega_f=0.5$, as shown in Eq.~\eqref{eq:eta_otto}.
Deformation moves the efficiency in both directions. For $p_f,q_f>1$ it exceeds the
Otto value, reaching $\eta=0.782$ at $(1.2,1.2)$; for $p_f,q_f<1$ it falls below it.
Equation~\eqref{eq:eta_general} expresses the result through the measurement-induced
energy changes. Relative to a uniformly scaled spectrum, changing the pairwise gaps
changes how much initial-Hamiltonian energy must be rejected for a given $\QM$.
The observed efficiencies follow this gap-weighted population redistribution. No
point on the evaluated grid with $\max(p_f,q_f)\le1$ exceeds the Otto value; this is
a numerical observation on the stated domain, not a general theorem for every
protocol or parameter choice.

The efficiency also depends on the asymmetry of the deformation, as
Fig.~\ref{fig:eta_contour} shows directly. If $\eta$ depended only on the symmetric
combination $\delta$, as it does at first order (Appendix~\ref{app:perturbative_expansion}), the Otto contour
$\eta=\eta_{\rm Otto}$ would coincide with the line $\delta=0$. Instead, the two lines
meet only at the undeformed point $(1,1)$. Away from it, the Otto contour bends below
the line $\delta=0$, so every asymmetric point on that line lies in the region
$\eta>\eta_{\rm Otto}$. For example, $(1.2,0.8)$ has $\delta=0$ but $\eta=0.534$. The
deformed engine can therefore outperform the undeformed oscillator even when the mean
deformation vanishes, an effect governed by the antisymmetric combination
$\gamma=(p_f-q_f)/2$. Because $[n]_{p,q}=[n]_{q,p}$, every quantity is even in
$\gamma$, so this dependence first appears at second order, consistent with
Appendix~\ref{app:perturbative_expansion}. It is nevertheless sizable: along the line
$\delta=0$ the efficiency increases as $\eta-\eta_{\rm Otto}\simeq0.8\,\gamma^2$,
taking the values $0.502$, $0.508$, $0.519$, and $0.534$ at $\gamma=0.05$, $0.10$,
$0.15$, and $0.20$, and the work output at $(1.2,0.8)$ exceeds the undeformed value by
$16\%$. The same holds away from $\delta=0$: the points $(0.7,1.2)$ and $(0.95,0.95)$
share $\delta=-0.05$, yet $\eta=0.465$ for the asymmetric point and $0.413$ for the
symmetric one. This fixed-mean comparison distinguishes the two-parameter family from the
standard slices $p=1$ with $q$ varying or
$p=q$. The condition $\delta=0$ selects only the undeformed oscillator, for which
$\eta=\eta_{\rm Otto}$ exactly by Eq.~\eqref{eq:eta_otto}. Independent control of the two coordinates allows the spectrum to be reshaped at
fixed mean deformation. A specially chosen one-dimensional path, for example
$p=1+\gamma$, $q=1-\gamma$, can also exhibit this effect; it is therefore not
a phenomenon excluded from all one-parameter parametrizations. Beyond perturbation
theory, the asymmetry also determines the physical regime: the large-$n$ spectrum
grows as $[n]_{p,q}\simeq\max(p,q)^n/|p-q|$, so the convergence criteria and the
self-adjointness of the measured quadrature in Sec.~\ref{sec:numerics} depend on
$\max(p,q)$ rather than on $\delta$. The points $(1,1)$ and $(1.2,0.8)$ share
$\delta=0$, yet the first describes an infinite oscillator and the second a
finite-level working medium. The high efficiency
does not conflict with the second law: the input is a nonthermal measurement channel,
and the cost of operating the measurement apparatus is not included in $\eta$.

After imposing the engine conditions of Eq.~\eqref{eq:engine_conditions}, the
physical working-medium efficiency satisfies
\begin{equation}
 0<\eta\lesssim0.78236,
 \label{eq:eta_range}
\end{equation}
where $\eta\to0$ continuously at the boundary $\Wcyc=0$ of the engine region and the
maximum occurs at $p_f=q_f=1.2$. The undeformed oscillator gives
$\eta_{\rm Otto}=0.5$, so deformation raises the efficiency by up to $56\%$ relative
to the Otto value.

Part of such an enhancement could, in principle, reflect only the larger energy scale
of the deformed final spectrum. To test selected resource constraints, we compare each deformed engine with
an undeformed oscillator that consumes the same resource, choosing the final
frequency $\omega_f'$ of the comparison oscillator accordingly. Because the undeformed
cycle has $\eta=1-\omega_i/\omega_f'$ exactly for any measurement strength,
Eq.~\eqref{eq:eta_otto}, and $\QM^{\rm HO}=\hbar^2/8m\sigma^2$ at the same physical
resolution, these comparisons are analytic. Table~\ref{tab:resource} lists three
choices: equal first-stroke work $W_I$, which is equivalent to equal mean final
energy since the initial state is fixed; equal lowest final gap
$E_1^{(f)}-E_0^{(f)}$; and equal measurement heat $\QM$, obtained by adjusting the
resolution of the comparison oscillator at $\omega_f'=\omega_f$. At the four tabulated points, the deformed engine is more efficient and delivers
more work under these selected constraints. In the equal-$\QM$ column the harmonic
reference uses $\sigma_{\rm HO}=\hbar/\sqrt{8m\QM}$, so its physical resolution differs
from that of the deformed system. Matching $\QM$ alone leaves the harmonic reference frequency free: its efficiency can approach unity as $\omega_f'/\omega_i$ increases. The equal-heat column therefore compares a chosen fixed-frequency reference, not the best harmonic engine under that sole constraint. No column matches every resource simultaneously,
and each comparison also changes the measured operator from a deformed quadrature
to a harmonic one. The cost of implementing that operator is not quantified. At $(1.2,1.2)$ the
efficiency exceeds that of the undeformed oscillator at equal $W_I$ by $28\%$, at
equal gap by $34\%$, and at equal $\QM$ by $56\%$, and the work output at equal $W_I$
is $3.8$ times larger. These selected comparisons indicate a role for relative level spacing; they do not
establish optimality over all harmonic reference protocols or equal detector costs. Equal resources also isolate the
asymmetry: $(1.1,1.1)$ and $(1.0,1.2)$ have the same $\delta$ and the same lowest gap,
yet the asymmetric point is more efficient, with $\eta=0.663$ compared with $0.654$.

\begin{table*}[tp]
\caption{Resource-normalized comparison at the parameters of
Table~\ref{tab:numerical_parameters}. For each deformed engine, $\omega_f'$ is the final
frequency of an undeformed oscillator with the same first-stroke work $W_I$ or the
same lowest final gap, and $\eta^{\rm HO}=1-\omega_i/\omega_f'$ is its exact
efficiency. At equal measurement heat the undeformed efficiency is $0.5$, using a changed resolution $\sigma_{\rm HO}=\hbar/\sqrt{8m\QM}$. Only the named resource is matched in each column.}
\label{tab:resource}
\begin{ruledtabular}
\begin{tabular}{ccccccc}
 & & \multicolumn{2}{c}{equal $W_I$} & \multicolumn{2}{c}{equal gap} & equal $\QM$ \\
$(p_f,q_f)$ & $\eta$ & $\omega_f'/\omega_i$ & $\eta^{\rm HO}$ & $\omega_f'/\omega_i$ & $\eta^{\rm HO}$ & $\eta^{\rm HO}$ \\
\hline
$(1.2,0.8)$ & 0.534 & 2.027 & 0.507 & 2.0 & 0.500 & 0.5 \\
$(1.1,1.1)$ & 0.654 & 2.255 & 0.557 & 2.2 & 0.545 & 0.5 \\
$(1.0,1.2)$ & 0.663 & 2.263 & 0.558 & 2.2 & 0.545 & 0.5 \\
$(1.2,1.2)$ & 0.782 & 2.562 & 0.610 & 2.4 & 0.583 & 0.5 \\
\end{tabular}
\end{ruledtabular}
\end{table*}

Finally, the two standard one-parameter $q$-oscillators are contained in the present
model as curves in the $(p,q)$ plane: the Arik--Coon oscillator, $[n]_q=(1-q^n)/(1-q)$,
corresponds to $p=1$, and the Biedenharn--Macfarlane oscillator,
$[n]_q=(q^n-q^{-n})/(q-q^{-1})$, to $pq=1$. Under the same protocol, the largest
efficiency along these curves within the retained domain is $0.663$, at $(1.0,1.2)$, for
the former and $0.556$, at $(1.2,1/1.2)$, for the latter, compared with $0.782$ for the
two-parameter oscillator at the same $\max(p,q)=1.2$. On both curves, $\delta=0$ only at
the undeformed point, so neither can reproduce the asymmetric enhancement discussed
above. The reported maximum at $p_f=q_f=1.2$ is also attained on the one-parameter diagonal family $p_f=q_f=s$; it is not inaccessible to all one-parameter models. This is a within-protocol comparison of spectral families, not a numerical
reproduction of Tong et al.~\cite{TONG2026131619}. Their published abstract describes
a selective measurement protocol, whereas ours discards the outcome. A direct
quantitative comparison with that engine requires its full protocol and remains
outside the present calculations.

\section{Conclusion and discussion}
\label{sec:conclusion}

We have formulated a measurement-driven quantum engine whose working medium is a
two-parameter $(p,q)$-deformed harmonic oscillator. The deformation changes the
relative spacing of the energy levels. It therefore modifies both the energy injected
by a nonselective Gaussian measurement and the work exchanged during the two
adiabatic strokes. With work defined as work done on the system, useful engine
operation corresponds to $\Wcyc<0$, and the working-medium efficiency is
$\eta=-\Wcyc/\QM$ for $\QM>0$.

The analytical results are as follows. We expressed the cycle energetics through the
measurement transition matrix $T_{nm}$ and evaluated it exactly, without quadrature,
by noting that the nonselective measurement acts as the dephasing superoperator
$\exp(-\mathcal{D}_x^2/8\sigma^2)$ of the deformed position quadrature. This yields
a finite-dimensional series for the measurement heat in powers of $\sigma^{-2}$, whose leading
term, Eq.~\eqref{eq:QM_weak}, shows that each level contributes in proportion to the
product of a deformed number and a deformed level spacing. For the undeformed
oscillator, this contribution is the same for every level, which recovers the known
state-independent heat $\hbar^2/8m\sigma^2$. We also showed that positivity of the
transition probabilities alone does not guarantee $\QM>0$; a sufficient condition is
that the measurement channel be unital and that the state entering the measurement be
passive with respect to the final Hamiltonian. Finally, we showed that the cycle
reproduces the Otto efficiency $1-\omega_i/\omega_f$ exactly whenever the spectrum is
scaled uniformly, whatever the source of the energy input. The Otto value is
therefore a property of the spectrum, not of the thermal bath, and it is the breaking
of uniform scaling by the deformation that allows $\eta$ to depart from it.

The infinite quadrature requires a domain specification: it is essentially
self-adjoint for $\max(p,q)\le1$, while above this range its self-adjoint extensions
are not unique. Nonuniqueness does not rule out an infinite oscillator after an
extension is chosen and the resulting energy traces are checked. Here we choose the
explicit finite model in Eq.~\eqref{eq:finite_boundary}. The leading weak-measurement
moment test is not a convergence theorem for its exact finite-strength measurement channel for the infinite oscillator.
At the revised parameters, $N=40$ to $60$ changes are small, but net work changes
slightly more than $10^{-3}$ relatively at the corner, and larger dimensions show
additional drift. These are finite-model sensitivity tests.

Within this domain, the cycle operates as an engine at $97.6\%$ of the parameter
points; it fails only in a small corner where both parameters are well below one.
Increasing the deformation raises the measurement heat by up to a factor of $2.9$ and
the work output by up to a factor of $4.6$ relative to the undeformed oscillator. The
working-medium efficiency lies in the range $0<\eta\lesssim0.78236$. It equals the Otto value
$0.5$ for the undeformed oscillator, exceeds it for $p_f,q_f>1$, and reaches its
maximum at $p_f=q_f=1.2$, an enhancement of $56\%$. The identity $\eta=1+\QT/\QM$ relates the gain to the energy rejected in
thermalization. At fixed mean deformation, varying $\gamma=(p_f-q_f)/2$ changes
performance at second and higher orders. This distinguishes the two-parameter
surface from the standard one-parameter slices compared here; it is not an
impossibility result for arbitrary one-dimensional deformation paths.

These conclusions should be read within the scope of the model. First, the
efficiency is a working-medium efficiency: the energy cost of implementing,
recording, and resetting the measurement apparatus is not included, and a complete
thermodynamic benchmark would require an explicit detector or autonomous measurement
model. The high efficiency does not conflict with the second law, because the input
is a nonthermal measurement channel. Second, for $\max(p_f,q_f)>1$ the reported results describe the specified finite
matrix model, not an extension-independent infinite oscillator. A physical realization
would need to justify its dimension, upper boundary, tunable level energies, and
measured quadrature. No such microscopic implementation is established here. Third,
the resource comparisons fix one selected constraint at a time. The equal-heat
comparison changes the harmonic reference resolution; fixed spectral width, identical
detector coupling, and control costs remain untested.

Future work should connect the deformation parameters to a concrete physical
implementation or to a controlled family of effective Hamiltonians, include
finite-time and nonadiabatic effects in realizations where the eigenbasis changes
during the work strokes, and account for the energetic cost of controlling the
deformation and of operating the measurement apparatus. These extensions would
further clarify the thermodynamic advantage of algebraic spectral engineering in
measurement-driven quantum engines.

\begin{acknowledgments}
This research was funded by Khalifa University of Science and Technology through Project ID KU-INT-RIG-2024-8474000739.
\end{acknowledgments}

\section*{Data Availability} 
The data that support the findings of this article are available upon reasonable request from the authors.

\appendix

\section{Analytical consistency checks}
\label{app:checks}

\subsection{Undeformed limit}

Setting $p,q\to 1$ gives $\pqnum{n}\to n$ and therefore
\begin{equation}
 E_n(\omega,1,1)=\frac{\hbar\omega}{2}(n+n+1)=\hbar\omega\left(n+\frac{1}{2}\right),
 \label{eq:undeformed_limit}
\end{equation}
which recovers the ordinary harmonic oscillator.

A useful benchmark for the measurement calculation is obtained in the undeformed final
oscillator, $p_f=q_f=1$. In the position representation, the nonselective Gaussian
channel in Eq.~\eqref{eq:measurement_operator} acts as
\begin{equation}
 \rho(x,x')\longmapsto \exp\left[-\frac{(x-x')^2}{8\sigma^2}\right]\rho(x,x'),
 \label{eq:app_gaussian_channel_kernel}
\end{equation}
which is equivalent to momentum diffusion with variance
\begin{equation}
 \Delta \pi^2=\frac{\hbar^2}{4\sigma^2}.
 \label{eq:app_momentum_diffusion}
\end{equation}
For the ordinary harmonic oscillator, this gives the state-independent measurement heat
\begin{equation}
 \QM^{\mathrm{HO}}=\frac{\hbar^2}{8m\sigma^2}.
 \label{eq:app_harmonic_measurement_benchmark}
\end{equation}
Since the transition matrix is evaluated exactly, Eq.~\eqref{eq:Texact}, the only
approximation in the numerical method is the Hilbert-space cutoff. The numerical
$\QM$ must therefore approach Eq.~\eqref{eq:app_harmonic_measurement_benchmark} as the
cutoff increases, and any deviation is a direct measure of the cutoff error. At the
parameters of Table~\ref{tab:numerical_parameters}, the method reproduces this value
in all displayed digits (Table~\ref{tab:convergence}).

\subsection{Diagonal deformation limit and finite truncation}
\label{subsec:finite_truncation}

When $p=q=s$, the quotient form of Eq.~\eqref{eq:pq_basic} is indeterminate, and the
deformed number takes the continuous limit
\begin{equation}
 [n]_{s,s}=n s^{n-1}.
 \label{eq:appendix_diagonal_limit}
\end{equation}
In the numerical calculations, this limit is obtained automatically because we use the
polynomial form of Eq.~\eqref{eq:pq_sum}, which is regular on and near the diagonal.

Parameters below unity require separate discussion~\cite{Boumali2017,Bonatsos1999}.
When $0<p,q<1$,
\begin{equation}
    [n]_{p,q}=\frac{p^{n}-q^{n}}{p-q}\to 0
\end{equation}
as $n\to\infty$, and consequently
\begin{equation}
    E_n=\frac{\hbar \omega}{2}\left([n]_{p,q}+[n+1]_{p,q}\right)\to 0 .
\end{equation}
The spectrum is then bounded and not monotonic: the energies first increase with $n$
and then decrease toward zero. A Gibbs state of such a Hamiltonian would not exist,
since the infinite partition function
\begin{equation}
 Z=\sum_{n=0}^{\infty}e^{-\beta E_n}
\label{eq:appendix_partition_function}
\end{equation}
diverges, its summand tending to one rather than zero. The truncated partition
function
\begin{equation}
    Z^{(N)}=\sum_{n=0}^{N-1}e^{-\beta E_n}
\end{equation}
is finite for every $N$, but it has no infinite-dimensional limit.

This divergence does not affect the present cycle, because the deformed Hamiltonian
is never thermalized. The only Gibbs state used is that of the initial, undeformed
Hamiltonian, and the final spectrum enters only through sums weighted by the initial
and post-measurement populations (Sec.~\ref{sec:numerics}). For $0<p_f,q_f<1$ the
final energies are bounded, so these sums converge rapidly, and the results in this
region are genuine infinite-oscillator results; at $(p_f,q_f)=(0.7,0.7)$, for example,
they are converged to better than $10^{-15}$ (Table~\ref{tab:convergence}). What the
non-monotonic spectrum does change is the sign guarantee for the measurement heat. The
spectral-ordering condition, Eq.~\eqref{eq:spectral_order_check}, fails, so the
passivity argument of Appendix~\ref{app:passivity_unitality} no longer applies, and $\QM$ must be evaluated directly. In the present calculations, it remains positive throughout the
domain.

Parameters above unity raise a different issue, which concerns the measured quadrature
rather than the spectrum. In the Fock basis, $\hat{x}_{p,q}$ is a Jacobi matrix with
zero diagonal and off-diagonal elements $b_n\propto\sqrt{[n+1]_{p,q}}$. From
Eq.~\eqref{eq:pq_basic} one finds the identity
\[
 [n]_{p,q}^2-[n-1]_{p,q}\,[n+1]_{p,q}=(pq)^{n-1},
\]
which also holds on the diagonal, where both sides equal $s^{2n-2}$. Hence
$b_{n-1}b_{n+1}\le b_n^2$ for every $n$. For $\max(p,q)\le1$, every term of
Eq.~\eqref{eq:pq_sum} is at most one, so $[n]_{p,q}\le n$ and
$\sum_n b_n^{-1}=\infty$; by Carleman's criterion, $\hat{x}_{p,q}$ is essentially
self-adjoint on finite linear combinations of Fock states. For $\max(p,q)>1$,
Eq.~\eqref{eq:pq_sum} gives $[n]_{p,q}\ge\max(p,q)^{n-1}$, so
$\sum_n b_n^{-1}<\infty$; together with $b_{n-1}b_{n+1}\le b_n^2$, Berezanskii's
theorem~\cite{Reiffenstein2025} then implies that $\hat{x}_{p,q}$ is not essentially self-adjoint and has a
one-parameter family of self-adjoint extensions. A Gaussian measurement of the
quadrature, Eq.~\eqref{eq:measurement_operator}, requires a self-adjoint operator, so
for $\max(p_f,q_f)>1$ it is defined only after a particular extension, or a particular
finite-dimensional truncation, has been chosen. The truncated deformed oscillator used
here is one such specification; its results depend weakly on $N$ over the range
reported in Table~\ref{tab:convergence} but need not converge as $N\to\infty$.

\subsection{First-law closure}

Combining Eqs.~\eqref{eq:work_I}, \eqref{eq:QM_general}, \eqref{eq:work_II}, and
\eqref{eq:thermalization_heat}, and defining $\Delta E_n=E_n^{(f)}-E_n^{(i)}$, gives
\begin{align}
 W_I+W_{II} &= \sum_n \Delta E_n\left(P_n^{(i)}-P_n^{(M)}\right),\label{eq:work_pair_appendix}\\
 \QM+\QT &= -\sum_n \Delta E_n\left(P_n^{(i)}-P_n^{(M)}\right).\label{eq:heat_pair_appendix}
\end{align}
Therefore,
\begin{equation}
 W_I+\QM+W_{II}+\QT=0.
 \label{eq:first_law_closure_appendix}
\end{equation}
This identity holds independently of the spectrum and the measurement, and provides a
compact analytical and numerical check of the adopted sign convention.

\section{Perturbative small-deformation expansion}
\label{app:perturbative_expansion}

\subsection{Expansion of the deformed spectrum}

Let
\begin{equation}
 p=1+a,\qquad q=1+b,
 \qquad \varepsilon=\max(|a|,|b|),
 \label{eq:pert_def_ab}
\end{equation}
with $|a|,|b|\ll 1$. The finite-sum representation in Eq.~\eqref{eq:pq_sum},
\begin{equation}
 \pqnum{n}=\sum_{r=0}^{n-1}p^{n-1-r}q^r,
 \label{eq:pert_finite_sum}
\end{equation}
is the natural starting point because it is polynomial in $p$ and $q$ and therefore remains regular on the diagonal $p=q$. To first order,
\begin{equation}
 p^{n-1-r}q^r
 \approx 1+(n-1-r)a+rb+\mathcal{O}(\varepsilon^2).
 \label{eq:pert_term_expansion}
\end{equation}
Using
\begin{equation}
 \sum_{r=0}^{n-1}(n-1-r)=\sum_{r=0}^{n-1}r=\frac{n(n-1)}{2},
 \label{eq:pert_sum_identity}
\end{equation}
one obtains
\begin{equation}
 \pqnum{n}\approx n+\delta\,n(n-1)+\mathcal{O}(\varepsilon^2),
 \quad
 \delta\equiv \frac{p+q}{2}-1=\frac{a+b}{2}.
 \label{eq:pert_basic_number}
\end{equation}
For $n=2$, Eq.~\eqref{eq:pert_basic_number} gives $2+2\delta=2+a+b$, which agrees exactly with $\pqnum{2}=p+q$ to this order. The antisymmetric deformation
\begin{equation}
 \gamma\equiv \frac{p-q}{2}=\frac{a-b}{2}
 \label{eq:pert_gamma}
\end{equation}
drops out at the first order. Thus, in the immediate neighborhood of $(p,q)=(1,1)$, the two-parameter deformation reduces perturbatively to the single symmetric control parameter $\delta$; the distinction between $p$ and $q$ enters only at second and higher orders.

\subsection{Energy spectrum to first order}

Substituting Eq.~\eqref{eq:pert_basic_number} into Eq.~\eqref{eq:energy_spectrum} gives
\begin{align}
 E_n(\omega,p,q)
 &\approx \frac{\hbar\omega}{2}\left[(2n+1)+\delta\,n\{(n-1)+(n+1)\}\right] \nonumber\\
 &=\hbar\omega\left(n+\frac{1}{2}\right)+\hbar\omega\,\delta\,n^2.
 \label{eq:pert_energy_expanded}
\end{align}
Equivalently,
\begin{equation}
E_n(\omega,p,q)\approx E_n^{(0)}(\omega)+\hbar\omega\,\delta\,n^2,
 \quad
 E_n^{(0)}(\omega)=\hbar\omega\left(n+\frac{1}{2}\right).
 \label{eq:pert_energy_spectrum}
\end{equation}
The leading deformation therefore acts as a Kerr-type quadratic correction to the harmonic spectrum, with strength determined by the average deviation $\delta=(p+q)/2-1$. The adjacent-level spacing becomes
\begin{equation}
 \Delta_n=E_{n+1}-E_n
 \approx \hbar\omega\left[1+\delta(2n+1)\right],
 \label{eq:pert_level_spacing}
\end{equation}
which confirms analytically that the deformed oscillator has excitation-dependent gaps. For $\delta>0$, the spacing grows with $n$, whereas for $\delta<0$ it decreases within the perturbative regime.

\subsection{First adiabatic work}

For the protocol used in the numerical study, the initial deformation is undeformed, $(p_i,q_i)=(1,1)$. The initial Gibbs probabilities are therefore those of the ordinary harmonic oscillator,
\begin{equation}
 P_n^{(i)}=\frac{e^{-\beta\hbar\omega_i(n+1/2)}}{Z_i},
 \qquad
 \bar n=\frac{1}{e^{\beta\hbar\omega_i}-1}.
 \label{eq:pert_initial_distribution}
\end{equation}
Using Eq.~\eqref{eq:pert_energy_spectrum} for $E_n^{(f)}$ and Eq.~\eqref{eq:work_I}, with
\begin{equation}
 \delta_f=\frac{p_f+q_f}{2}-1,
 \label{eq:pert_delta_f}
\end{equation}
yields
\begin{equation}
 W_I\approx \hbar(\omega_f-\omega_i)\left(\bar n+\frac{1}{2}\right)
 +\hbar\omega_f\,\delta_f\,\langle n^2\rangle_i.
 \label{eq:pert_work_I_moment}
\end{equation}
The geometric thermal distribution satisfies
\begin{equation}
 \langle n\rangle_i=\bar n,
 \qquad
 \langle n^2\rangle_i=2\bar n^2+\bar n,
 \label{eq:pert_thermal_moments}
\end{equation}
so the first adiabatic work has the closed perturbative form
\begin{equation}
 W_I\approx \hbar(\omega_f-\omega_i)\left(\bar n+\frac{1}{2}\right)
 +\hbar\omega_f\,\delta_f\left(2\bar n^2+\bar n\right).
 \label{eq:pert_work_I}
\end{equation}
The first term is the ordinary undeformed contribution. The second term is the leading deformation correction and scales linearly with $\delta_f$ and quadratically with the thermal occupation at high temperature. For $\delta_f>0$, the compression-like stroke therefore requires additional work input, consistent with the deformation-induced lifting of the final spectrum.

\subsection{Second adiabatic work and net-work criterion}

After the measurement stroke, the final-basis populations $P_n^{(M)}$ need not be thermal. Define their moments by
\begin{equation}
 \langle n^k\rangle_M\equiv \sum_{n=0}^{N-1}n^k P_n^{(M)}.
 \label{eq:pert_measurement_moments}
\end{equation}
Using Eq.~\eqref{eq:pert_energy_spectrum} in Eq.~\eqref{eq:work_II} gives
\begin{equation}
 W_{II}\approx -\hbar(\omega_f-\omega_i)\left(\langle n\rangle_M+\frac{1}{2}\right)
 -\hbar\omega_f\,\delta_f\,\langle n^2\rangle_M.
 \label{eq:pert_work_II}
\end{equation}
Combining Eqs.~\eqref{eq:pert_work_I} and \eqref{eq:pert_work_II},
\begin{align}
 \Wcyc
 &\approx \hbar(\omega_f-\omega_i)\left(\bar n-\langle n\rangle_M\right) \nonumber\\
 &\quad +\hbar\omega_f\,\delta_f\left[(2\bar n^2+\bar n)-\langle n^2\rangle_M\right].
 \label{eq:pert_cycle_work}
\end{align}
Equation~\eqref{eq:pert_cycle_work} is the perturbative counterpart of the numerical engine boundary. For a compression stroke $\omega_f>\omega_i$ with $\delta_f>0$, negative net work requires the measurement-induced redistribution to increase the relevant first and/or second occupation moments enough that the negative return-stroke work overcomes the positive work input of the first adiabatic stroke. For other signs of $\delta_f$, the same equation gives the corresponding competition between the harmonic contribution and the deformation-dependent second-moment contribution.

\subsection{Measurement heat: leading operator structure}

A closed first-order expression for $\QM$ requires the first-order expansion of the Gaussian Kraus matrix elements, because the measured quadrature itself depends on the deformed ladder operators. The operator structure can nevertheless be displayed transparently on the diagonal $p=q=s=1+\delta$. From Eq.~\eqref{eq:diagonal_limit},
\begin{equation}
 [n]_{s,s}=n s^{n-1}\approx n\left[1+(n-1)\delta\right],
 \label{eq:pert_diagonal_basic}
\end{equation}
and hence
\begin{equation}
 \sqrt{[n]_{s,s}}\approx \sqrt{n}\left[1+\frac{\delta}{2}(n-1)\right].
 \label{eq:pert_ladder_matrix_element}
\end{equation}
Let $a_0$ and $a_0^{\dagger}$ denote the undeformed annihilation and creation operators and $N$ the ordinary number operator. Since $Na_0=a_0(N-1)$, Eq.~\eqref{eq:pert_ladder_matrix_element} gives
\begin{equation}
 a_{s,s}\approx a_0+\frac{\delta}{2}\,a_0(N-1),
 \qquad
 a_{s,s}^{\dagger}\approx a_0^{\dagger}+\frac{\delta}{2}(N-1)a_0^{\dagger}.
 \label{eq:pert_ladder_operator}
\end{equation}
Therefore, the diagonal-deformation quadrature is
\begin{equation}
 \hat{x}_{s,s}\approx \hat{x}_0+\frac{\delta}{2}\sqrt{\frac{\hbar}{2m\omega_f}}
 \left[a_0(N-1)+(N-1)a_0^{\dagger}\right].
 \label{eq:pert_position_operator}
\end{equation}
The measured position quadrature thus acquires a number-weighted correction. This explains why the response of $\QM$ to deformation is strongest for transitions involving higher occupied levels: the measurement samples energy gaps and quadrature matrix elements whose leading corrections grow with occupation number. A closed expression for the leading measurement heat, valid for arbitrary deformation, is given by the weak-measurement result of Eq.~\eqref{eq:QM_weak}; the perturbative spectral and work formulas above do not depend on it.

\section{Passivity and unitality guarantee for \texorpdfstring{$\QM\geq 0$}{QM >= 0}}
\label{app:passivity_unitality}

This appendix gives an explicit finite-dimensional proof of the sufficient condition used in Sec.~\ref{subsec:measurement_stroke}: a passive state acted on by the present nonselective Gaussian measurement channel cannot yield negative measurement heat.

\subsection{Transition-matrix identities}

\textit{Lemma 1.} The transition matrix of the Hermitian Gaussian measurement channel is symmetric:
\begin{equation}
 T_{nm}=T_{mn}.
 \label{eq:app_T_symmetry}
\end{equation}
Indeed, $M_\alpha=M_\alpha^{\dagger}$ implies
\begin{equation}
 \langle n;f|M_\alpha|m;f\rangle
 =\langle m;f|M_\alpha|n;f\rangle^{*},
 \label{eq:app_hermitian_matrix_elements}
\end{equation}
and therefore
\begin{equation}
 \left|\langle n;f|M_\alpha|m;f\rangle\right|^2
 =\left|\langle m;f|M_\alpha|n;f\rangle\right|^2.
 \label{eq:app_modulus_symmetry}
\end{equation}
Integrating over $\alpha$ in Eq.~\eqref{eq:transition_matrix} proves Eq.~\eqref{eq:app_T_symmetry}.

\textit{Lemma 2.} The nonselective measurement channel
\begin{equation}
 \Phi(\rho)=\int_{-\infty}^{\infty}M_\alpha\rho M_\alpha\,\dd\alpha
 \label{eq:app_channel}
\end{equation}
is unital, $\Phi(\Id)=\Id$, and its transition matrix is doubly stochastic. The completeness relation in Eq.~\eqref{eq:measurement_completeness} gives trace preservation and hence the column sums
\begin{equation}
 \sum_n T_{nm}
 =\int_{-\infty}^{\infty}\langle m;f|M_\alpha^{\dagger}M_\alpha|m;f\rangle\,\dd\alpha
 =1.
 \label{eq:app_column_stochastic}
\end{equation}
Since the Kraus operators are Hermitian, the same completeness relation also gives
\begin{equation}
 \Phi(\Id)=\int_{-\infty}^{\infty}M_\alpha M_\alpha^{\dagger}\,\dd\alpha
 =\int_{-\infty}^{\infty}M_\alpha^{\dagger}M_\alpha\,\dd\alpha
 =\Id.
 \label{eq:app_unitality}
\end{equation}
Taking the $(n,n)$ matrix element of Eq.~\eqref{eq:app_unitality} gives the row sums
\begin{equation}
 \sum_m T_{nm}
 =\int_{-\infty}^{\infty}\langle n;f|M_\alpha M_\alpha^{\dagger}|n;f\rangle\,\dd\alpha
 =1.
 \label{eq:app_row_stochastic}
\end{equation}
Thus $T_{nm}$ is doubly stochastic; by Lemma 1, it is also symmetric. These identities hold independently of the values of $p_f$, $q_f$, and $\sigma$, and they hold exactly in the retained $N$-level space when $T_{nm}$ is evaluated from Eq.~\eqref{eq:Texact}, as confirmed by the diagnostics in Eq.~\eqref{eq:norm_sym_error}.

\subsection{Nonnegativity of measurement heat}

\textit{Proposition.} Suppose the retained final spectrum is ordered as
\begin{equation}
 E_0^{(f)}\leq E_1^{(f)}\leq E_2^{(f)}\leq\cdots,
 \label{eq:app_ordered_spectrum}
\end{equation}
and the state entering the measurement stroke is passive with respect to $H_f$, namely
\begin{equation}
 n<m\quad \Rightarrow \quad P_n^{(i)}\geq P_m^{(i)}.
 \label{eq:app_passive_populations}
\end{equation}
Then $\QM\geq 0$.

Starting from the symmetrized expression in Eq.~\eqref{eq:QM_symmetrized},
\begin{equation}
 \QM=\frac{1}{2}\sum_{n,m}
 \left(E_n^{(f)}-E_m^{(f)}\right)T_{nm}
 \left(P_m^{(i)}-P_n^{(i)}\right),
 \label{eq:app_QM_symmetrized}
\end{equation}
the summand is invariant under $n\leftrightarrow m$ because both the energy difference and the population difference change sign, while $T_{nm}=T_{mn}$. Pairing the two terms $(n,m)$ and $(m,n)$ gives
\begin{equation}
 \QM=\sum_{n<m}
 \left(E_n^{(f)}-E_m^{(f)}\right)T_{nm}
 \left(P_m^{(i)}-P_n^{(i)}\right).
 \label{eq:app_QM_pair_sum}
\end{equation}
For each pair $n<m$, spectral ordering gives $E_n^{(f)}-E_m^{(f)}\leq 0$, passivity gives $P_m^{(i)}-P_n^{(i)}\leq 0$, and $T_{nm}\geq 0$ because it is an integral over a modulus squared. Hence, every pair contribution is nonnegative:
\begin{equation}
 \QM=\sum_{n<m}
 \underbrace{\left(E_n^{(f)}-E_m^{(f)}\right)}_{\leq 0}
 \underbrace{T_{nm}}_{\geq 0}
 \underbrace{\left(P_m^{(i)}-P_n^{(i)}\right)}_{\leq 0}
 \geq 0.
 \label{eq:app_QM_positive}
\end{equation}
This proves the stated sufficient condition.

Equality, $\QM=0$, occurs only if every pair $n<m$ satisfies at least one of the following conditions: $T_{nm}=0$, $E_n^{(f)}=E_m^{(f)}$, or $P_n^{(i)}=P_m^{(i)}$. In words, the measurement must connect no pair of levels that are simultaneously nondegenerate in energy and nondegenerate in incoming population.

\subsection{Relation to the general passive-state theorem}

The preceding proof is the finite-dimensional specialization of the general result that unital operations cannot lower the energy of a passive state \cite{Pusz1978CMP,Lenard1978JSP}. In the final-energy basis, Lemmas 1 and 2 show that a symmetric, doubly stochastic matrix represents the measurement-induced population map. By the Birkhoff--von Neumann theorem, this population map can be viewed as a convex mixture of permutations of the energy eigenbasis. Equation~\eqref{eq:app_QM_positive} is the corresponding pairwise proof for the present Gaussian measurement channel and makes explicit why passivity and spectral ordering are sufficient to guarantee $\QM\geq 0$.

\section{Dependence on the final frequency and the measurement resolution}
\label{app:scans}

\begin{figure*}[tp]
 \centering
 \includegraphics[width=\textwidth]{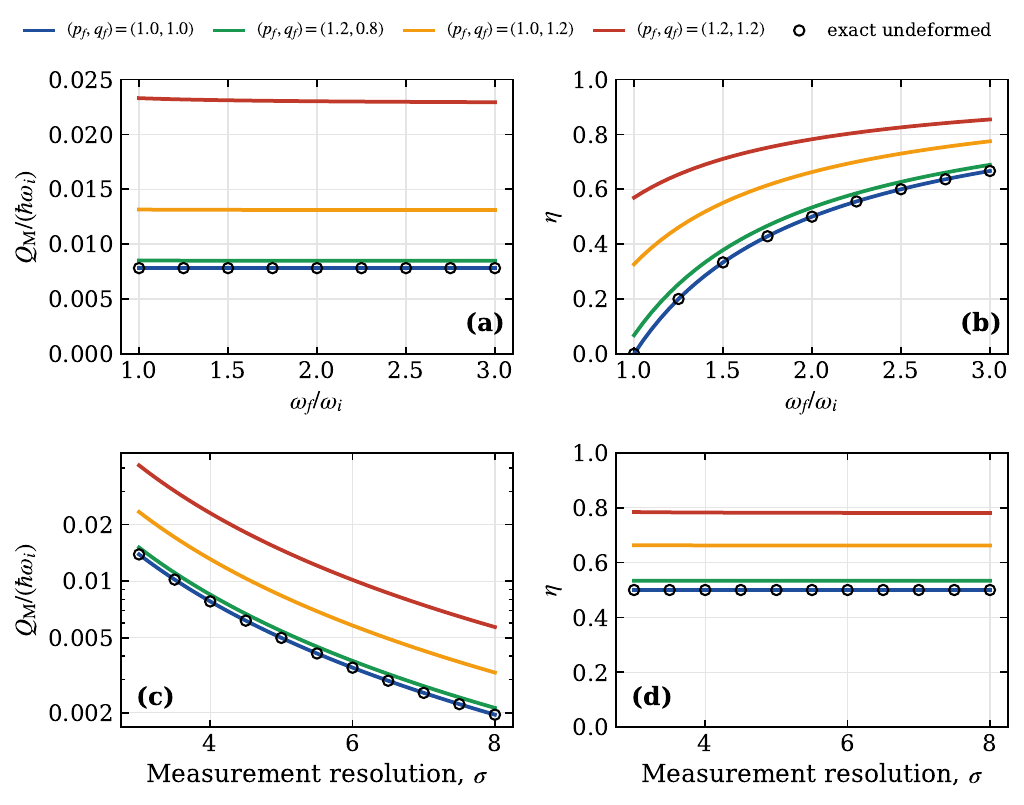}
 \caption{Measurement heat and working-medium efficiency for four final deformations,
$(p_f,q_f)=(1.0,1.0)$, $(1.2,0.8)$, $(1.0,1.2)$, and $(1.2,1.2)$, at
$\beta\hbar\omega_i=1$ and $N=40$. (a) $\QM$ and (b) $\eta$ as functions of the
compression ratio $\omega_f/\omega_i$ at $\sigma=4$; (c) $\QM$ (logarithmic scale)
and (d) $\eta$ as functions of the measurement resolution $\sigma$ at
$\omega_f/\omega_i=2$. Open circles show the exact undeformed results,
$\QM^{\rm HO}=\hbar^2/8m\sigma^2$ and $\eta_{\rm Otto}=1-\omega_i/\omega_f$. The
relative change of $\QM$ on increasing the dimension from $N=40$ to $N=60$ is below
$3\times10^{-6}$ for the first three deformations over both scans, and at most
$7\times10^{-3}$ for $(1.2,1.2)$, at the ends of the scans where the measurement is
effectively strongest ($\omega_f=\omega_i$ and $\sigma=3$).}
 \label{fig:app_scans}
\end{figure*}

Figure~\ref{fig:app_scans} shows how the measurement heat and the working-medium efficiency
depend on the compression ratio $\omega_f/\omega_i$, at $\sigma=4$, and on the
measurement resolution $\sigma$, at $\omega_f/\omega_i=2$, for four representative
deformations. All other parameters are those of Table~\ref{tab:numerical_parameters}.
For the undeformed oscillator, the numerical results coincide with the exact values
$\QM^{\rm HO}=\hbar^2/8m\sigma^2$ and $\eta_{\rm Otto}=1-\omega_i/\omega_f$ (open
circles) throughout both scans.

The measurement heat is almost independent of the final frequency
[Fig.~\ref{fig:app_scans}(a)]. This follows from the weak-measurement result,
Eq.~\eqref{eq:QM_weak}. The prefactor scales as $\omega_f^{-1}$, whereas every level
spacing $\Delta_n^{(f)}$, and therefore every $B_n$, scales as $\omega_f$, while the
populations $P_n^{(i)}$ do not depend on $\omega_f$. The two factors cancel, so at
leading order in $\sigma^{-2}$ the energy injected by a measurement of fixed physical
resolution is independent of the compression. The deformation, by contrast, changes
$B_n$ and raises $\QM$ at every frequency. The small residual decrease with
$\omega_f$ at $(1.2,1.2)$ reflects the higher-order terms of Eq.~\eqref{eq:QM_series}:
the oscillator length decreases with $\omega_f$, so the dimensionless resolution
$\sigmat$ of Eq.~\eqref{eq:dimensionless_sigma} increases and the measurement becomes
effectively weaker.

The efficiency, Fig.~\ref{fig:app_scans}(b), follows $\eta_{\rm Otto}=1-\omega_i/\omega_f$
for the undeformed oscillator and lies above it for all three deformed cases. The
advantage is largest without compression. At $\omega_f=\omega_i$ the undeformed engine
produces no work, since the spectrum does not change, whereas the deformed engines
reach $\eta\simeq0.07$, $0.33$, and $0.57$ for $(1.2,0.8)$, $(1.0,1.2)$, and
$(1.2,1.2)$, respectively. The deformation of the spectrum alone therefore suffices to
convert measurement backaction into work. As the compression increases, the
frequency change dominates the spectral change, and the curves approach the Otto
value. The point $(1.2,0.8)$, which has vanishing mean deformation $\delta=0$, remains
above the Otto curve over the entire range, confirming that the effect of the
antisymmetric deformation identified in Sec.~\ref{subsec:efficiency_results} does not
depend on the compression ratio.

The dependence on the measurement resolution is shown in
Figs.~\ref{fig:app_scans}(c) and (d). The measurement heat decreases approximately as
$\sigma^{-2}$ for all deformations, as expected from Eq.~\eqref{eq:QM_weak}, and the
deformed curves lie above the undeformed one by a factor that is nearly independent of
$\sigma$. The efficiency, in contrast, is essentially independent of $\sigma$. At
leading order the population change $\delta P_n$ produced by the measurement is
proportional to $\sigma^{-2}$, and the exact expression of Eq.~\eqref{eq:eta_general}
is a ratio of two sums that are both linear in $\delta P_n$, so the dependence on
$\sigma$ cancels. At leading weak-measurement order, the resolution controls the injected energy
while canceling from the efficiency ratio. Higher-order corrections can make
the efficiency resolution-dependent; the spectrum alone does not determine it exactly.

\end{document}